\documentclass[%
 aip,
 amsmath,amssymb,
 reprint,%
]{revtex4-2}

\usepackage[usenames,dvips]{color}

\usepackage{graphicx}
\usepackage{dcolumn}
\usepackage{bm}

\usepackage[utf8]{inputenc}
\usepackage[T1]{fontenc}
\usepackage{mathptmx}
\usepackage{etoolbox}

\usepackage{mathtools}
\usepackage{epstopdf}
\usepackage{esint}
\usepackage{xcolor}
\usepackage{dsfont}
\usepackage{nicefrac}

\usepackage[colorlinks=true,citecolor=magenta,linkcolor=magenta, urlcolor=magenta]{hyperref}

\newcommand{\bea}{\begin{eqnarray}}
\newcommand{\eea}{\end{eqnarray}}
\newcommand{\bt}{\textbf}

\newcommand{\phd}{\phantom{\dag}}
\newcommand{\ph}{\phantom{.}}

\newcommand{\noi}{\noindent}
\newcommand{\no}{\nonumber}

\makeatletter
\def\@email#1#2{%
 \endgroup
 \patchcmd{\titleblock@produce}
  {\frontmatter@RRAPformat}
  {\frontmatter@RRAPformat{\produce@RRAP{*#1\href{mailto:#2}{#2}}}\frontmatter@RRAPformat}
  {}{}
}%
\makeatother

\begin{document}

\preprint{AIP/123-QED}

\title{Green Function Invariants for Floquet Topological Superconductivity\\Induced by Proximity Effects}

\author{Mohamed Assili}
\author{Panagiotis Kotetes}
\email{kotetes@baqis.ac.cn}
\affiliation{Beijing Academy of Quantum Information Sciences, Beijing 100193, China}


\begin{abstract}
We bring forward a Green function approach for the prediction of Floquet to\-po\-lo\-gi\-cal phases in driven superconductor-semiconductor hybrids. Although it is common to treat the superconducting component as a mere Cooper-pair reservoir, it was recently pointed out that such an appro\-xi\-ma\-tion breaks down in the presence of driving, due to the emergence of level broadening. Here, we go beyond these recent works and prescribe how to construct the Floquet topological invariants for such driven hybrids. Specifically, we propose to first obtain the midgap quasi-energy spectra by including the hermitian part of the semiconductor's self-energy and, subsequently, read out the respective level broadenings by projecting the anti-hermitian part of the self-energy onto the quasi-energy eigenvectors. We exemplify our approach for a Rashba nanowire coupled to a superconductor and a time-dependent Zeeman field. Using our method, we obtain the Floquet band structure, the respective level broadenings, and  the topological invariants. Our analysis reinforces the need to properly account for the self-energy, and corroborates that broadening effects can hinder the observation of the Floquet topological phases and especially of those harboring Majorana $\pi$ modes.
\end{abstract}

\maketitle

\section{Introduction}

Hybrid superconductor-semiconductor quantum devices have recently emerged as unique playgrounds for en\-gi\-nee\-ring exotic quasi-particles, which fea\-ture pro\-per\-ties that ra\-di\-cal\-ly differ from those of electrons~\cite{JaySau,AliceaPRB,LutchynPRL,OregPRL}. For instance, these particle-like excitations may carry fractional spin or electric charge, and they may even sa\-ti\-sfy anyonic exchange statistics, i.e., they behave neither as fermions nor as bosons. Majorana zero modes (MZMs) stand out as prominent excitations of this type, since they constitute charge-neutral Ising anyons, and open perspectives for {\color{black}fault-tolerant} quantum com\-pu\-ting~\cite{Nayak2008,Ivanov2001,Kitaev2003}. While a number of {\color{black}cutting-edge} experiments have claimed the creation and manipulation of MZMs in superconductor-semiconductor hybrids~\cite{
Mourik,MTEarly,Das,MT,Sven,Sole,Azure}, the possible presence of MZMs in these systems still remains uncertain and continues to be under scrutiny~\cite{PradaRev,Valentini}.

Although experimental efforts have so far exclusively focused on static Majorana platforms, numerous theo\-re\-ti\-cal proposals have explored the possible emergence of Floquet topological superconductivity in these hybrids. In fact, one finds a long list of studies con\-cer\-ning, among others, Rashba nanowires~\cite{Reynoso,WMLiuHet,klinovaja2016,Thakurathi2017,LiuLevchenkoLutchyn,YangZhesen,Forcellini,Mondal,RoyBasu} and planar Josephson junctions~\cite{ChangnanPeng,Liu2019}. Aside from the enhanced degree of tu\-na\-bi\-li\-ty provided by the drive, such systems also exhibit pro\-per\-ties that are inaccessible to their static analogs~\cite{OkaRev,RudnerRev}. The most stri\-king dif\-fe\-ren\-ce is the emergence of an additional species of quasi-particles, those termed as Majorana $\pi$ modes (MPMs), which behave {\color{black}si\-mi\-lar\-ly} to MZMs but appear at a nonzero energy set by the drive~\cite{Jiang,WMLiu,DELiu2013}. Notably, the simultaneous presence of MZMs and MPMs in the same hybrid unlocks alternative schemes for imple\-men\-ting quantum computing~\cite{BomantaraPRL,bomantara2018,BelaBauer}.

In spite of the fact that there exist numerous studies on Floquet topological superconductivity, the vast majority of these works have assumed oversimplified descriptions for the superconducting proximity effect. As it is customary, in such hybrids we are primarily interested in the pro\-per\-ties of the semicon\-duc\-ting component. The superconductor functions instead as a reservoir which sets the temperature, chemical potential, and {\color{black}Cooper-pair} density of the semiconductor~\cite{Antipov,Woods,Mikkelsen,Reeg}. In this context, the simplification that is routinely carried out is to neglect the frequency dependence of the self-energy induced in the semiconductor due to its coupling to the superconductor~\cite{McMillan,SauProxi,PotterComparison}. While this is a legitimate approximation when we are inte\-re\-sted in the low-energy properties of the hybrid, this approach ge\-ne\-ral\-ly breaks down when driving is added.

\begin{figure}[t!]
\begin{center}
\includegraphics[width=0.92\columnwidth]{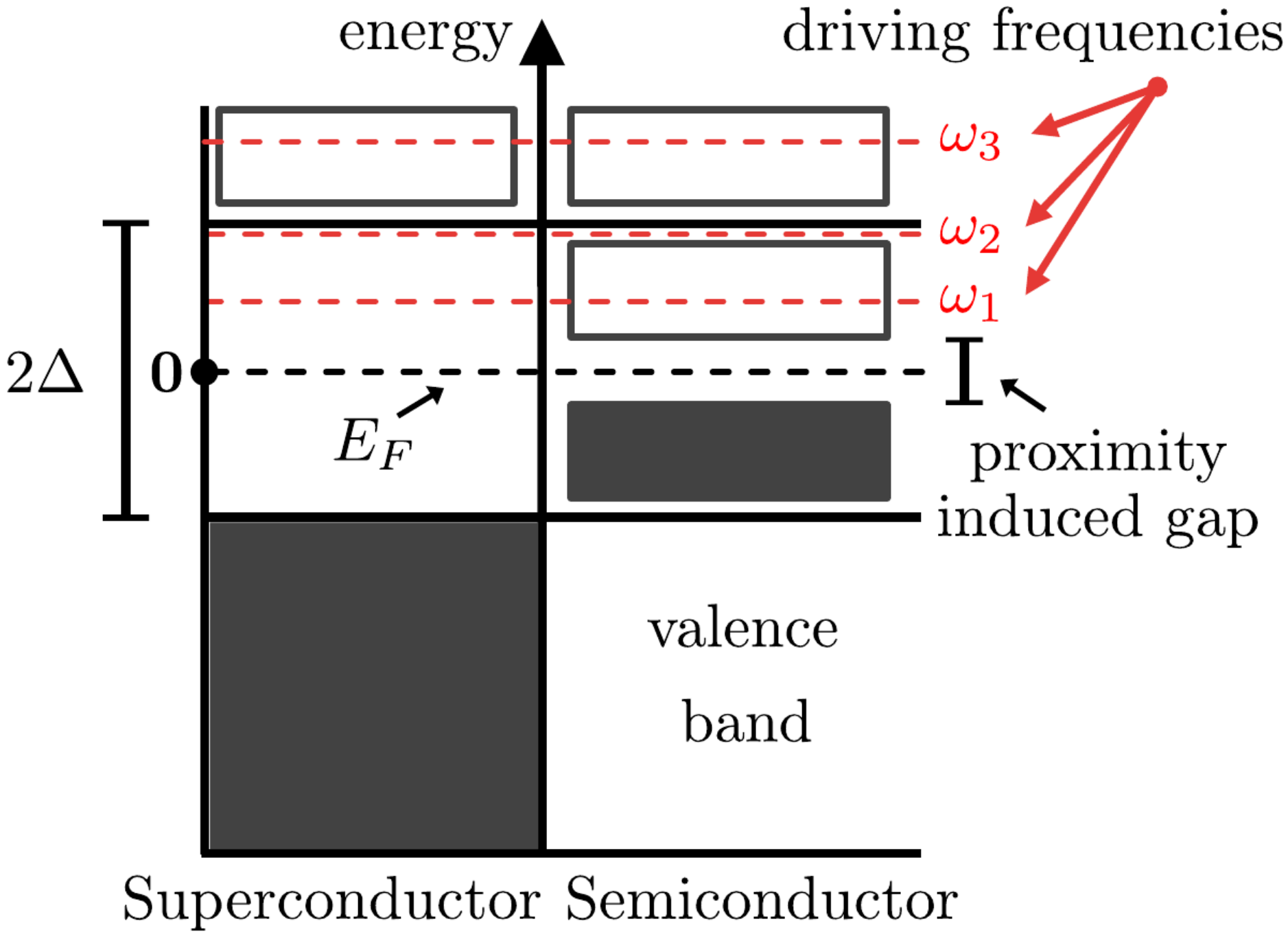}
\end{center}
\caption{{\color{black}Band diagram of the superconductor-semiconductor hybrid structure after the two materials get in contact and equilibrium is established. Occupied (empty) bands are denoted with shaded (empty) boxes. The Fermi energy of the superconductor also sets the Fermi level of the entire hybrid system, since it is here considered to constitute a good metal and, therefore, it can be viewed as a particle bath. Additionally, the pro\-xi\-mi\-ty to the superconductor induces a pairing gap in the conduction band of the semiconductor, which is of focus in this work. The behavior of the driven system depends on the hier\-archy of the frequency of the periodic driving $\omega$ and the superconducting gap $\Delta\geq0$. For frequency values $\omega_1\ll\Delta$ and $\omega_3\gg\Delta$, the energy level broadening is negligible. In stark contrast, for intermediate frequencies $\omega_2\sim\Delta$, the broadening becomes substantial and predominantly hinders the emergence of Majorana $\pi$ modes. Note that the downfolding to the first Floquet zone is not considered here.}}
\label{fig:Figure1}
\end{figure}

This crucial aspect was recently pointed out in Refs.~\cite{LiuLevchenkoLutchyn,YangZhesen,Forcellini}. In more detail, Ref.~\onlinecite{YangZhesen} emphasized that discarding the ener\-gy dependence of the self-energy is justified only when the dri\-ving frequency is much smaller than the pai\-ring gap of the superconductor. In contrast, in all other frequency regimes the driving unavoidably couples the semiconductor to predominantly metallic-like bands of the superconducting bath. {\color{black}See Fig.~\ref{fig:Figure1} for a sketch.} Hence, aside from the desired proximity-induced pairing gap, unwanted level broadening and dissipation are also introduced to the semiconductor~\cite{LiuLevchenkoLutchyn}. While dissipation can serve as a means to reach a Floquet steady state, the appearance of broadening generally obscures and masks the to\-po\-lo\-gi\-cal properties of the hybrid. Previous works studied the local density of states and the degree of observabi\-li\-ty of MZMs and MPMs~\cite{YangZhesen,Forcellini}, thus confirming the failure of this routine appro\-xi\-ma\-tion in driven hybrids. In spite of these great advances, the important question of how to obtain the topological phase diagrams for such hybrids {\color{black}remains unaddressed}.

In this work, we resolve this pressing issue by making use of the Green function formalism to properly define the quasi-energy operator (QEO). The QEO describes the dynamics of these driven hybrids in the frequency domain and allows to properly incorporate the self-energy of the semiconductor due to the proximity effect. With the QEO at hand, we proceed to define the respective topological invariants which govern the engineered Floquet topological superconducting phases~\cite{Obuse,Asboth2012,AsbothObuse,Tarasinski,Asboth2014,Yao2017,Kennes,Assili2024}. As we discuss, the QEO is obtained by retrieving only the hermitian (principal) part of the retarded Green function of the semiconductor. As we demonstrate, the anti-hermitian part of it can be subsequently employed to evaluate the broadening of each eigenvector corresponding to a given quasi-energy level by means of a sui\-table projection.

In the remainder, we first analyze the general problem of interest by con\-si\-de\-ring the coupled time-dependent Schr\"odinger equations for the semiconductor and superconductor components, while later on we make a connection to the Green function for\-ma\-lism and how to employ it to obtain the QEO. Although our approach is kept general, we also derive concrete expressions for the self-energy for commonly-assumed physical sce\-na\-rios, such as, wideband superconductor-semiconductor tunnel coupling in the pre\-sen\-ce of single-harmonic driving. We conclude by exemplifying our approach in the case of a single-channel one-dimensional Rashba nanowire in proximity to a conventional spin-singlet superconductor. The nanowire is additionally under the influence of a Zeeman field which consists of both static and time-dependent components oriented along the axis of the nanowire. We determine the quasi-energy band structure and topological invariants characterizing the hybrid for various driving-frequency regimes. To infer the observability of the ari\-sing topological superconducting phases, we further eva\-lua\-te the level broadenings of the quasi-energy bands.

\section{Modeling Driven Hybrid Devices}

We begin by laying out the fundamental equations that describe the systems of interest. In the absence of driving, the semiconductor and the superconductor are assumed to be governed by the continuum and translationally-invariant Hamiltonians $H_{\rm sm}(\bm{p})$ and $H_{\rm sc}(\bm{p},\bm{q})$, respectively. These two systems are addi\-tio\-nal\-ly coupled to each other by means of a spatially-local tunnel-coupling~\cite{McMillan,SauProxi,PotterComparison},  which is determined by the term ${\rm T}(\bm{p},\bm{q})$. It is important to emphasize that here  $H_{\rm sc}(\bm{p},\bm{q})$ defines a mean-field Hamiltonian for the superconductor, in which we have split the momenta {\color{black}labeling} it using two distinct vectors, i.e., $\bm{p}$ and $\bm{q}$. This is because, in such hybrids, the semiconductor component typically constitutes a lower-dimensional system compared to the superconductor. The momentum vector $\bm{p}$ incorporates all the momenta which characterize the semiconductor, while $\bm{q}$ includes all the additional momenta which need to be introduced in order to describe the superconductor, as a result of their dimensional mismatch.

In the remainder, we consider the additional presence of periodic driving terms $V_{\rm sm}(t,\bm{p})$ and $V_{\rm sc}(t,\bm{p})$. Putting all these ingredients together leads to the following two coupled time-dependent Schr\"odinger equations (where we set $\hbar=1$):
\begin{align}
&\big[i\partial_t-H_{\rm sm}(\bm{p})-V_{\rm sm}(t,\bm{p})\big]\bm{\Psi}(t,\bm{p})=\int d\bm{q}\,{\rm T}(\bm{p},\bm{q})\bm{\Phi}(t,\bm{p},\bm{q}),\\
&\big[i\partial_t-H_{\rm sc}(\bm{p},\bm{q})-V_{\rm sc}(t,\bm{p},\bm{q})\big]\bm{\Phi}(t,\bm{p},\bm{q})={\rm T}^\dag(\bm{p},\bm{q})\bm{\Psi}(t,\bm{p}).
\end{align}

\noi The Schr\"odinger equation for the semiconducting segment is formulated in terms of the state vector $
\bm{\Psi}(t,\bm{p})=\big[\psi_{e,\uparrow}(t,\bm{p}),\,\psi_{e,\downarrow}(t,\bm{p}),\,\psi_{h,\downarrow}(t,-\bm{p}),\,-\psi_{h,\uparrow}(t,-\bm{p})\big]^\intercal$, where $^\intercal$ indicates matrix transposition. In a similar fa\-shion, for the superconductor we employ the following state vector $\bm{\Phi}(t,\bm{p},\bm{q})=$ $\big[\phi_{e,\uparrow}(t,\bm{p},\bm{q}),\,\phi_{e,\downarrow}(t,\bm{p},\bm{q}),\,\phi_{h,\downarrow}(t,-\bm{p},-\bm{q}),\,$ $-\phi_{h,\uparrow}(t,-\bm{p},-\bm{q})\big]^\intercal$. Given these definitions for the state vectors, we  express any Hamiltonian element as a Kronecker product of the form $\tau_\mu\otimes\sigma_\nu$, where $\mu,\nu=0,1,2,3$. In the present context, $\tau_{1,2,3}$ ($\sigma_{1,2,3}$) denote Nambu (spin) Pauli matrices, while $\tau_0$ ($\sigma_0$) corresponds to their associated identity matrix. From now on, we omit the Kronecker product symbol and the identity matrices.

In view of an improved perturbative approach, it appears particularly useful to perform a unitary transformation which eliminates the driving terms $V_{\rm sm}(t,\bm{p})$ and $V_{\rm sc}(t,\bm{p},\bm{q})$.~\cite{Benito2014,LiuLevchenkoLutchyn} We consider the unitary transformations
$\bm{\Psi}(t,\bm{p})\mapsto{S}_{\rm sm}(t,\bm{p};t_0)\bm{\Psi}(t,\bm{p})$ and $\bm{\Phi}(t,\bm{p},\bm{q})\mapsto{S}_{\rm sc}(t,\bm{p},\bm{q};t_0)\bm{\Phi}(t,\bm{p},\bm{q})$, where we introduced:
\bea
{S}_{\rm sm}(t,\bm{p};t_0)&=&e^{-i\int^t_{t_0}d\tau\,V_{\rm sm}(\tau,\bm{p})}\,,
\label{eq:UnitarySM}\\
{S}_{\rm sc}(t,\bm{p},\bm{q};t_0)&=&e^{-i\int^t_{t_0}d\tau\,V_{\rm sc}(\tau,\bm{p},\bm{q})}\,.
\label{eq:UnitarySC}
\eea

\noi The above manipulations allow us to re-express the respective Schr\"odinger equations compactly according to:
\bea
\big[i\partial_t-H_{\rm sm}(t,\bm{p})\big]\bm{\Psi}(t,\bm{p})=\int d\bm{q}\,{\rm T}(t,\bm{p},\bm{q})\bm{\Phi}(t,\bm{p},\bm{q}),\quad\\
\big[i\partial_t-H_{\rm sc}(t,\bm{p},\bm{q})\big]\bm{\Phi}(t,\bm{p},\bm{q})={\rm T}^\dag(t,\bm{p},\bm{q})\bm{\Psi}(t,\bm{p}),\quad
\eea

\noi where we made use of the definitions:
\bea
H_{\rm sm}(t,\bm{p})&=&{S}_{\rm sm}^\dag(t,\bm{p};t_0)H_{\rm sm}(\bm{p}){S}_{\rm sm}(t,\bm{p};t_0)\,,
\label{eq:RotatedHSM}\\
H_{\rm sc}(t,\bm{p},\bm{q})&=&{S}_{\rm sc}^\dag(t,\bm{p},\bm{q};t_0)H_{\rm sc}(\bm{p},\bm{q}){S}_{\rm sc}(t,\bm{p},\bm{q};t_0)\,,\quad\\
{\rm T}(t,\bm{p},\bm{q})&=&{S}_{\rm sm}^\dag(t,\bm{p};t_0)
{\rm T}(\bm{p},\bm{q}){S}_{\rm sc}(t,\bm{p},\bm{q};t_0)\,.\label{eq:TFourier}
\eea

\noi In spite of the seemingly complex form of the above terms, in the case of a single-harmonic periodic driving the above quantities end {\color{black}up taking} a rather simple form that {\color{black} renders the treatment of the time-dependent problem more transparent. Even more, it can also allow for approximate analytical solutions.}

\section{Floquet Topological Invariants}

The {\color{black}throughout assumed} time periodicity of the dri\-ving terms with a period $T$, which is expressed as $V_{\rm sm}(t,\bm{p})=V_{\rm sm}(t+T,\bm{p})$ and $V_{\rm sc}(t,\bm{p},\bm{q})=V_{\rm sc}(t+T,\bm{p},\bm{q})$ allows us to make use of the Floquet theorem~\cite{OkaRev,RudnerRev}. According to this, the two state vectors can be expressed as $
\bm{\Psi}_\varepsilon(t,\bm{p})=e^{-i\varepsilon t}\bm{u}_\varepsilon(t,\bm{p})$ and $
\bm{\Phi}_\varepsilon(t,\bm{p},\bm{q})=e^{-i\varepsilon t}\bm{v}_\varepsilon(t,\bm{p},\bm{q})$, where $\varepsilon$  corresponds to the so-called quasi-energy of the system. This variable lives in the {\color{black}first} Floquet zone, which is identified with the interval $[-\omega/2,\omega/2)$, where we introduced the driving frequency $\omega=2\pi/T$. These two newly-introduced state vectors are time-periodic and, thus, satisfy the relations $\bm{u}_\varepsilon(t,\bm{p})=\bm{u}_\varepsilon(t+T,\bm{p})$ and $\bm{v}_\varepsilon(t,\bm{p},\bm{q})=\bm{v}_\varepsilon(t+T,\bm{p},\bm{q})$. By virtue of their periodicity, these state vectors can be expanded in terms of Fourier series, i.e.,
$\bm{u}_\varepsilon(t,\bm{p})=\sum_ne^{in\omega t}\bm{u}_{\varepsilon;n}(\bm{p})$ and $\bm{v}_\varepsilon(t,\bm{p},\bm{q})=\sum_ne^{in\omega t}\bm{v}_{\varepsilon;n}(\bm{p},\bm{q})$ with $n\in\mathbb{Z}$. By exploiting the above properties, we can now reformulate the two coupled Schr\"odinger equations as follows:
\begin{widetext}
\bea
\sum_m\big[(\varepsilon-n\omega)\delta_{n,m}-H_{{\rm sm};n-m}(\bm{p})\big]\bm{u}_{\varepsilon;m}(\bm{p})&=&\sum_m\int d\bm{q}\,{\rm T}_{n-m}(\bm{p},\bm{q})\bm{v}_{\varepsilon;m}(\bm{p},\bm{q}),\\
\sum_m\big[(\varepsilon-n\omega)\delta_{n,m}-H_{{\rm sc};n-m}(\bm{p},\bm{q})\big]\bm{v}_{\varepsilon;m}(\bm{p},\bm{q})&=&\sum_m{\rm T}_{m-n}^\dag(\bm{p},\bm{q})\bm{u}_{\varepsilon;m}(\bm{p}),
\eea
\end{widetext}

\noi where $\delta_{n,m}$ denotes the Kronecker delta in Floquet space. The above expressions were obtained by means of re-expressing $H_{\rm sm}(t,\bm{p})$, $H_{\rm sc}(t,\bm{p},\bm{q})$, and ${\rm T}(t,\bm{p},\bm{q})$ in terms of Fourier series with respective coefficients $H_{{\rm sm};n}(\bm{p})$, $H_{{\rm sc};n}(\bm{p},\bm{q})$, and ${\rm T}_n(\bm{p},\bm{q})$. These follow from the following definition of the Fourier series:
\bea
f_n=\frac{1}{T}\int_0^Tdt\,e^{-in\omega t}\,f(t)\,.
\label{eq:FourierSeries}
\eea

By employing the two coupled equations one can now identify the quasi-energies for the entire system. However, since we are mainly interested in the effect of the superconductor onto the semiconductor, i.e., the so-called superconducting proximity effect, it is helpful to eliminate the super\-con\-duc\-ting degrees of freedom and obtain a nonlinear quasi-energy equation for the semiconductor alone. To facilitate this procedure, we introduce the inverse Floquet-Green function~\cite{Martinez} for the superconductor, whose matrix elements in Floquet space are given by the following relation:
\begin{align}
G_{{\rm sc};nm}^{-1}(\varepsilon,\bm{p},\bm{q})=(\varepsilon-n\omega)\delta_{n,m}-H_{{\rm sc};n-m}(\bm{p},\bm{q})\,.
\end{align}

\noi With the help of the above, we can express the state vector characterizing the superconductor according to:
\begin{align}
\bm{v}_{\varepsilon;\ell}(\bm{p},\bm{q})=\sum_{s,m} G_{{\rm sc};\ell s}(\varepsilon,\bm{p},\bm{q}){\rm T}_{m-s}^\dag(\bm{p},\bm{q})\bm{u}_{\varepsilon;m}(\bm{p})\,.
\end{align}

\noi This allows us to end up with an effective nonlinear Schr\"odinger equation which dictates the dynamics of the semiconductor and reads as:
\begin{align}
\sum_m{\cal G}_{{\rm sm};nm}^{-1}(\varepsilon,\bm{p})\bm{u}_{\varepsilon;m}(\bm{p})=\bm{0}.
\label{eq:SemiDynamics}
\end{align}

\noi The above equation is expressed in terms of the inverse of the dressed Green function of the semiconductor:
\begin{align}
{\cal G}_{{\rm sm};nm}^{-1}(\varepsilon,\bm{p})=G_{{\rm sm};nm}^{-1}(\varepsilon,\bm{p})-\Sigma_{{\rm sm};nm}(\varepsilon,\bm{p})\,,
\end{align}

\noi which takes contributions from the bare Floquet-Green function of the semiconductor:
\begin{align}
G_{{\rm sm};nm}^{-1}(\varepsilon,\bm{p})=(\varepsilon-n\omega)\delta_{n,m}-H_{{\rm sm};n-m}(\bm{p})\,,
\end{align}

\noi and the self-energy induced due to the proximity effect:
\bea
&&\Sigma_{{\rm sm};nm}(\varepsilon,\bm{p})=\int d\bm{q} \sum_{\ell,s}{\rm T}_{n-\ell}(\bm{p},\bm{q})\qquad\qquad\qquad\no\\
&&\qquad\qquad\qquad\quad\qquad \times G_{{\rm sc};\ell s}(\varepsilon,\bm{p},\bm{q}){\rm T}_{m-s}^\dag(\bm{p},\bm{q}).\qquad
\label{eq:Self-Energy}
\eea

It is important here to mention that the naive eva\-lua\-tion of the self-energy matrix elements does not ne\-ces\-sa\-ri\-ly lead to hermitian terms. As a result, it appears not straightforward to determine the precise form of the QEO which controls the quasi-energy band structure and can be employed to define the Floquet topological invariants. When the above expression of the self-energy is fully hermitian, the matrix ele\-ments ${\cal H}_{{\rm sm};nm}(\varepsilon,\bm{p})$ of the QEO are immediately obtained from those of the opposite inverse dressed Floquet-Green function, i.e., ${\cal H}_{{\rm sm};nm}(\varepsilon,\bm{p})=-{\cal G}_{{\rm sm};nm}^{-1}(\varepsilon,\bm{p})$. The respective Floquet topological invariants at the re\-le\-vant $\varepsilon=0$ and $\varepsilon=\pi/T=\omega/2\equiv-\omega/2$ quasi-energies are obtained by correspondingly fi\-xing the quasi-energy value in the QEO to be $\varepsilon=0$ and $\varepsilon=\omega/2$. There exist prior works in which Floquet topological invariants where eva\-lua\-ted in the pre\-sen\-ce of hermitian self-energy matrix elements, such as, in Floquet topological Anderson insulators~\cite{Titum2015,Titum2017}.

To our knowledge, the approach concerning how to obtain the Floquet to\-po\-lo\-gi\-cal invariants in systems where the naive evaluation of the self-energy may lead to nonhermitian results has not yet been addressed. In fact, a previous work in hybrid systems {\color{black}has} focused on the retarded and advanced self-energies~\cite{YangZhesen}, which are obtained using Eq.~\eqref{eq:Self-Energy} by following the prescription below:
\begin{align}
\Sigma^{R,A}_{{\rm sm};nm}(\varepsilon,\bm{p})=\Sigma_{{\rm sm};nm}(\varepsilon\pm i0^+,\bm{p}).
\end{align}

\noi Knowing the retarded or the advanced self-energy enables us to obtain the spectral function, which simultaneously encodes information about the quasi-energy dispersions and their corresponding broadening~\cite{Mahan}. Based on the analytical properties of the retarded and advanced Green functions, it is straightforward to associate the band structure and broa\-de\-ning with their hermitian [$\widetilde{\Sigma}_{{\rm sm};nm}(\varepsilon,\bm{p})$] and anti-hermitian [$\Gamma_{{\rm sm};nm}(\varepsilon,\bm{p})$] components, which we respectively define as:
\bea
\widetilde{\Sigma}_{{\rm sm};nm}(\varepsilon,\bm{p})&=&\frac{\Sigma^A_{{\rm sm};nm}(\varepsilon,\bm{p})+\Sigma^R_{{\rm sm};nm}(\varepsilon,\bm{p})}{2}\,,\quad
\label{eq:NiceSelf}\\
\Gamma_{{\rm sm};nm}(\varepsilon,\bm{p})&=&\frac{\Sigma^A_{{\rm sm};nm}(\varepsilon,\bm{p})-\Sigma^R_{{\rm sm};nm}(\varepsilon,\bm{p})}{2i}\,.
\label{eq:Broadening}
\eea

\noi The broadening part does not determine the quasi-energy band structure and, therefore, it cannot affect the ari\-sing topological phase transitions and Floquet topological invariants. However, the broadening is decisive in restric\-ting the obser\-va\-bi\-li\-ty of the emerging topological phases.

Based on the above arguments, we are now in a position to return to the present case of interest, and define the matrix elements of the QEO as follows:
\bea
{\cal H}_{{\rm sm};nm}(\varepsilon,\bm{p})&=&-\widetilde{{\cal G}}^{-1}_{{\rm sm};nm}(\varepsilon,\bm{p})\no\\
&\equiv&-G^{-1}_{{\rm sm};nm}(\varepsilon,\bm{p})+\widetilde{\Sigma}_{{\rm sm};nm}(\varepsilon,\bm{p}),\quad
\label{eq:QuasiEnergyOP}
\eea

\noi where the self-energy $\widetilde{\Sigma}_{{\rm sm};nm}(\varepsilon,\bm{p})$ can be either obtained from Eq.~\eqref{eq:NiceSelf} after evaluating the retarded or advanced self-energy, or, directly from Eq.~\eqref{eq:Self-Energy} after taking the principal value (${\cal P}$) of the integral, i.e.:
\begin{align}
\int d\bm{q}\phd\mapsto\phd{\cal P}\int d\bm{q}\,.
\end{align}

The above prescription is one of the central results of this work. In fact, employing such an approach in order to obtain the band structure of systems with a self-energy is known already from other types of problems, which do not include the periodic driving. For example, such a recipe has been discussed for impurity problems, where an isolated impurity is coupled to a band of conduction electrons~\cite{Mahan}. There, when the retarded self-energy of the impurity is real, the impurity constitutes a true bound state whose energy is affected by the conduction electrons. In contrast, when the retarded self-energy is purely imaginary the impurity acts as a scattering re\-so\-nan\-ce without having its energy modified. Therefore, the self-energy which affects the spectrum arises from the real (hermitian) part of the retarded Green function, which is obtained by choosing the principal value of similar integrals appearing in that analysis. As it is discussed in detail in Ref.~\onlinecite{Mahan} and other sources mentioned therein, choosing the type of self-energy to be employed corresponds to fixing the boundary condition for the impurity problem, that is, whether one is interested in studying standing, ingoing, or, outgoing waves. The principal value prescription picks out standing wave solutions, which allows us to infer the band structure of the system.

The problem that we investigate in this manuscript is not dif\-fe\-rent to the above impurity problem. The semiconductor can be viewed as a sort of a more complex impurity which is embedded in the bath of Cooper pairs and observes a self-energy due to the superconductor. When interested in de\-ri\-ving the (quasi-energy) band structure of the semiconductor we need to consider the principal value of the arising integrals in order to guarantee that the self-energy is hermitian.

\section{Self-Energy in Typical Situations}\label{sec:Typical}

In this section, we further explore the structure of the self-energy in cases where certain typical conditions are satisfied. Specifically, we examine various assumptions which become relevant when describing these hybrid systems, and obtain an expression for the self-energy.

\subsection{Superconductor and Tunnel-Coupling Hamiltonian Terms}

Our first assumption is to consider the routinely used model for a conventional superconductor within mean-field theory. This model is inva\-riant under spin rotations and translations, and takes the form:
\begin{align}
H_{\rm sc}(\bm{p},\bm{q})=\left(\frac{\bm{p}^2+\bm{q}^2}{2m_0}-E_F\right)\tau_3+\Delta\tau_1\,,
\label{eq:Sc-H}
\end{align}

\noi where we have formally split the momenta into the parts $\bm{p}$ and $\bm{q}$, while we introduced the bare electron mass $m_0$. In the cases of interest, the superconductor is an ideal conductor in the normal phase. Therefore, $E_F\gg\Delta$, since the Fermi energy $E_F$ is of the order of a few eVs, while the pai\-ring gap $\Delta$ is at best case scenarios about a couple of meVs. Typically, we are interested in the properties of the semiconductor near a particular minimum or maximum of the band structure. Hence, we can safely assume that $|\bm{p}|^2/2m_0\ll E_F$ holds, which in turn implies that $\bm{p}$ can be dropped from the Hamiltonian in Eq.~\eqref{eq:Sc-H} when evaluating the self-energy of Eq.~\eqref{eq:Self-Energy}. Under this assumption, the remaining dispersion $\xi(\bm{q})=q^2/2m_0-E_F$, where $q=|\bm{q}|$, can be linearized near the Fermi momenta $q_F$ which are obtained from $q_F^2=2m_0E_F$. Another common approximation concerns the tunnel-coupling element ${\rm T}(\bm{p},\bm{q})$, which is typically considered within the wideband approximation, i.e., the dependence on the various momenta is dropped. Thus, ${\rm T}(\bm{p},\bm{q})\simeq {\rm T}\,\tau_3$, where $\tau_3$ is the electric charge operator.

\subsection{Typical Time-Periodic Drives}

We now discuss common scenarios regarding the drives. Typically, it is considered that the drive influen\-ces only the semiconductor, which is usually the material component of primary interest, while the superconductor is assumed completely undriven or experiencing negligible driving. This implies that $V_{\rm sc}(t,\bm{p})$ can be discarded. Moreover, in most cases, the drive is considered to consist of a single harmonic, that is, it takes the form:
\begin{align}
V_{\rm sm}(t,\bm{p})=\beta_{\rm sm}(\bm{p})\omega\cos(\omega t+\zeta)\,.
 \end{align}

\noi While here ${S}_{\rm sc}(t,\bm{p},\bm{q};t_0)=\mathds{1}$, we find that for the semiconductor the respective unitary operator takes the form:
\begin{align}
{S}_{\rm sm}(t,\bm{p};t_0)
%
=e^{-i\beta_{\rm sm}(\bm{p})[\sin(\omega t+\zeta)-\sin(\omega t_0+\zeta)]}\,.
\end{align}

\noi In the remainder, we proceed by choosing $t_0=-\zeta/\omega$ which yields the simplified expression:
\begin{align}
{S}_{\rm sm}(t,\bm{p};t_0=-\zeta/\omega)=e^{-i\beta_{\rm sm}(\bm{p})\sin(\omega t+\zeta)}\,.
\label{eq:UnitarySMsimple}
\end{align}

\subsection{Form of the Self-Energy} \label{sec:SelfieForm}

Under all the above conditions, it is straightforward to obtain the time-dependent tunnel-coupling Hamiltonian ele\-ment ${\rm T}(t,\bm{p},\bm{q})$, which takes the form {\color{black}${\rm T}(t,\bm{p},\bm{q})=e^{i\beta_{\rm sm}(\bm{p})\sin(\omega t+\zeta)}\,{\rm T}\,\tau_3$.} By means of Eq.~\eqref{eq:TFourier} it leads to the Fourier series coefficients shown below:
{\color{black}
\begin{align}
{\rm T}_n(\bm{p},\bm{q})
%
=e^{i\zeta n}J_n[\beta_{\rm sm}(\bm{p})]\,{\rm T}\,\tau_3,
\end{align}
}

\noi where $J_n$ denotes the Bessel function of the first kind of
order $n\in\mathbb{Z}$. Straightforward manipulations detailed in Appendix~\ref{app:AppendixA} lead to the following expressions:
\begin{widetext}
\bea
-\widetilde{\Sigma}_{{\rm sm};nm}(\varepsilon,\bm{p})
&=&\Gamma e^{i\zeta(n-m)}\sum_\ell J_{n-\ell}[\beta_{\rm sm}(\bm{p})]\frac{\varepsilon-\ell\omega-\Delta\tau_1}
{\sqrt{\Delta^2-(\varepsilon-\ell\omega)^2}}J_{m-\ell}^\dag[\beta_{\rm sm}(\bm{p})]
\,\Theta\Big[\Delta^2-(\varepsilon-\ell\omega)^2\Big]\,,\label{eq:RealSelfieSimple}\\
\Gamma_{{\rm sm};nm}(\varepsilon,\bm{p})&=&
\Gamma e^{i\zeta(n-m)}\sum_\ell J_{n-\ell}[\beta_{\rm sm}(\bm{p})]
\frac{\varepsilon-\ell\omega-\Delta\tau_1}
{\sqrt{(\varepsilon-\ell\omega)^2-\Delta^2}}J_{m-\ell}^\dag[\beta_{\rm sm}(\bm{p})]\,\Theta\Big[(\varepsilon-\ell\omega)^2-\Delta^2\Big]{\rm sgn}(\varepsilon-\ell\omega),\qquad
\label{eq:BroadeningSimple}
\eea
\end{widetext}

\noi where we introduced $\Gamma=\pi\nu_F{\rm T}^2$. {\color{black} Here, $\nu_F$ is the density of states of the superconductor, which is evaluated at the Fermi level  $E_F$ of its normal phase. Note that, as we verify in Appendix~\ref{app:AppendixBB}, when the driving amplitude goes to zero, we consistently recover the static self-energy from the above equations.}

{\color{black}With the help of the above expressions, one obtains the QEO in Eq.~\eqref{eq:QuasiEnergyOP} and, in turn, the quasi-energy bands $\varepsilon_\nu(\bm{p})$, which are labeled by the band index $\nu$. Subsequently, one projects the broadening matrix onto the associated band eigenvector in order to infer the respective band broa\-de\-ning. It is crucial to mention that the non-negative nature of the dia\-go\-nal elements $\Gamma_{{\rm sm};nn}(\varepsilon,\bm{p})$, which are the only ones that are nonzero when the drive is absent, also ensures that the projection of $\Gamma_{{\rm sm};nm}(\varepsilon,\bm{p})$ onto a quasi-energy eigenvector always yields a non-negative value for the effective broadening.}

\subsection{High-Frequency Expansion}\label{sec:HFexpansion}

Quite often, the driving frequency exceeds the band width of the undriven system, and allows us to obtain a simplified {\color{black}description} of the system. In this limit, the system behaves in a quasi-static manner, with topological phases of insulators being solely associated with gap clo\-sings and reopenings at zero quasi-energy, while no topologically protected $\pi$-modes can emerge. In this section, we explore the physical scenario in which $|\varepsilon|<\Delta\ll\omega$. Due to the given hierarchy of the three energy scales, only the physics of the $n=0$ Floquet mode becomes re\-le\-vant for the topological properties of the system. Thus, the goal here is to obtain an {\color{black}approximate QEO and, in turn, an effective} Hamiltonian for the $n=0$ Floquet mode by integrating out all the other Floquet modes which happen to reside at high energies.

{\color{black}As we detail in Appendix~\ref{app:AppendixDD}, in the simultaneous high-frequency and low-driving-amplitude regime,} {\color{black} we find that the effective inverse Green function for the zero-th Floquet mode in the semiconductor becomes:
\bea
&&\big[{\cal G}^R_{{\rm sm};00}(\varepsilon,\bm{p})\big]^{-1}\approx \varepsilon-H_{{\rm sm};0}(\bm{p})+\Gamma\frac{\varepsilon-\Delta\tau_1}{\sqrt{\Delta^2-\varepsilon^2}}\no\\
&&+\sum_{m\neq0}\frac{m\omega}{(m\omega)^2+\Gamma^2}\,\Lambda_{{\rm sm};0m}(\varepsilon,\bm{p})\Lambda_{{\rm sm};m0}(\varepsilon,\bm{p})\no\\
&&+\sum_{m\neq0}\frac{i\Gamma}{(m\omega)^2+\Gamma^2}\,\Lambda_{{\rm sm};0m}(\varepsilon,\bm{p})\Lambda_{{\rm sm};m0}(\varepsilon,\bm{p}),\quad
\label{eq:HighFreqFreenRet}
\eea

\noi where we introduced the auxiliary quantities $\Lambda_{{\rm sm};nm}(\varepsilon,\bm{p})=H_{{\rm sm};n-m}(\bm{p})+\widetilde{\Sigma}_{{\rm sm};nm}(\varepsilon,\bm{p})
$. These are nonzero only for $n\neq m$.

In the above, we encounter the three following distinct contributions. The first row contains the usual inverse Green function for the semiconductor when no driving is present and $|\varepsilon|<\Delta$.~\cite{SauProxi,PotterComparison} The term in the second row corresponds to the usual high-frequency corrections due to the driving~\cite{eckardt2015}, pro\-per\-ly extended to include the broa\-dening and the off-diagonal self-energy terms. Note that for $\Gamma\rightarrow0$ one indeed retains the usual high-frequency expansion term. Lastly, in the third row one finds the broa\-de\-ning obtained due to the driving-induced mixing of the zero-th Floquet mode with the metallic bands of the superconductor. Notably, in contrast to the odd in $m\leftrightarrow-m$ prefactor appearing in the second row, the third row contains a prefactor which is even in  $m\leftrightarrow-m$. This dif\-fe\-ren\-ce is crucial in order to ensure that the resulting Green function is of the retarded type. At the same time, this guarantees the non-negative nature of the broadening term which is enforced thanks to the relation $\Lambda_{{\rm sm};nm}^\dag(\varepsilon,\bm{p})\equiv \Lambda_{{\rm sm};mn}(\varepsilon,\bm{p})$.
}

\section{Application to a Hybrid Nanowire}\label{sec:Application}

In this section, we apply the above formalism to a pro\-to\-ty\-pi\-cal hybrid nanowire system~\cite{AliceaPRB,LutchynPRL,OregPRL} in the presence of a driven external magnetic field{\color{black}, which can arise from the magnet of the cryostat. We investigate the topolo\-gi\-cal pro\-per\-ties of the semiconducting} segment by pro\-per\-ly incorpo\-ra\-ting the self-energy arising due to the superconducting pro\-xi\-mi\-ty effect.

\subsection{Model Hamiltonian}

\begin{figure}[t!]
\begin{center}
\includegraphics[width=0.91\columnwidth]{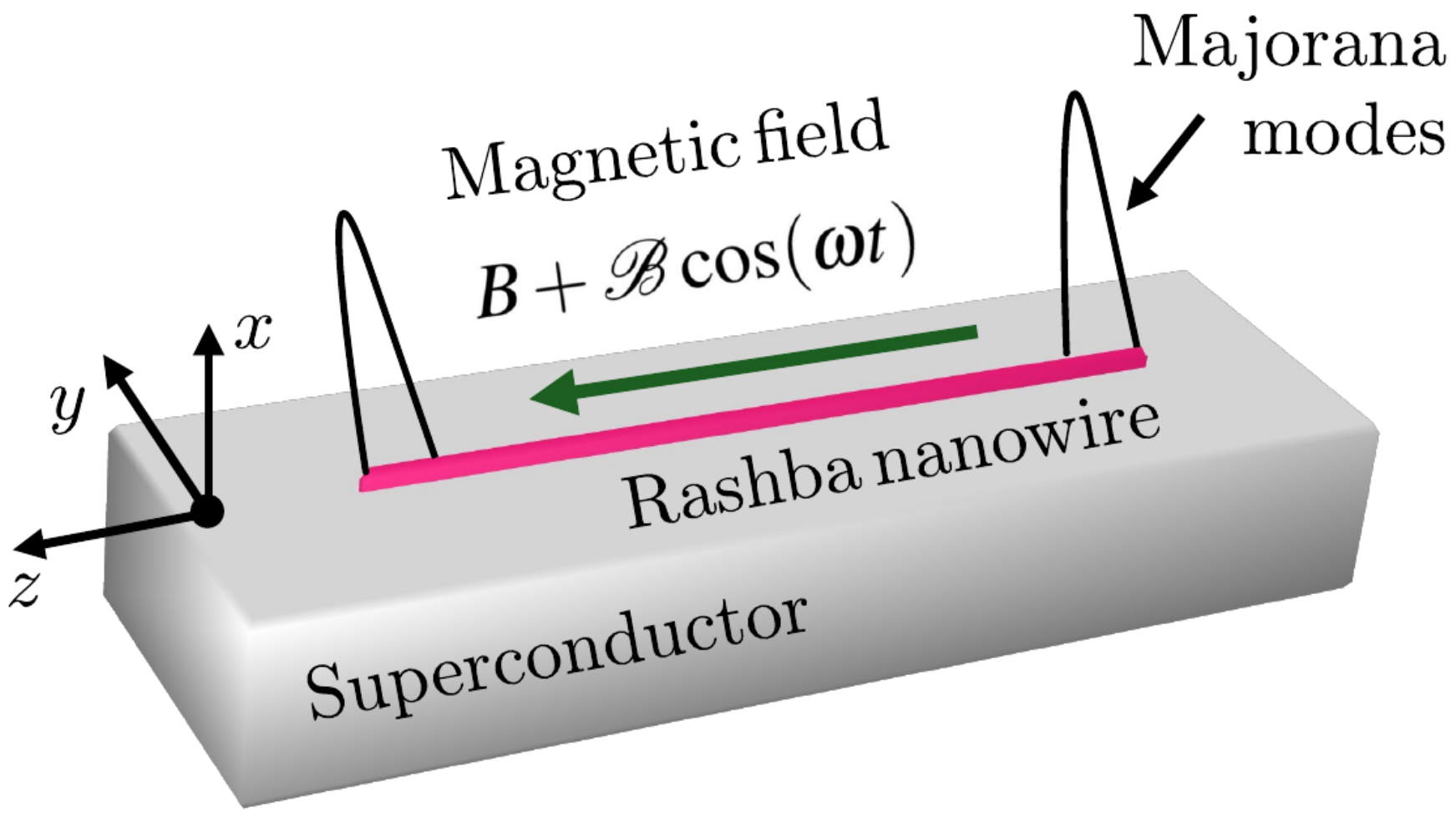}
\end{center}
\caption{{\color{black}Schematic illustration of the concrete driven superconductor-semiconductor hybrid investigated in this work. The semiconductor component consists of an ideal one-dimensional nanowire which is coupled to a conventional superconductor that plays the role of the bath. The nanowire is simultaneously subject to a time-periodic magnetic field $B+{\cal B}\cos(\omega t)$ that is oriented along the nanowire's axis. Majorana zero and $\pi$ edge modes emerge at the nanowire ends.}}
\label{fig:Figure2}
\end{figure}

We consider a hybrid system which consists of a semiconducting nanowire in contact with a bulk superconductor, whose superconducting gap is $\Delta\geq0$. A magnetic field $\bm{B}=B\hat{\bm{z}}$ is applied along the $z$ axis, which coincides with the axis of the nanowire. {\color{black}See Fig.~\ref{fig:Figure2} for a schematic depiction of the proposed experimental setup.} The Hamiltonian of the superconductor is given by Eq.~\eqref{eq:Sc-H}, while the one which describes the semiconducting part is written as follows:
\begin{eqnarray}
H_{\rm sm}(p_z)=\left(\frac{p_z^2}{2m_{\ast}}-\mu \right)\tau_3+E_Z\sigma_3
+\alpha p_z\tau_3\sigma_2,
\label{eq:HamiltonianExample}
\end{eqnarray}

\noi where $\alpha$ is the strength of the
Rashba spin-orbit coupling (SOC), $\mu$ is the chemical potential, $E_Z=g\mu_BB/2$ defines the static Zeeman energy, $\mu_B$ is the Bohr magneton, $g$ is the Land\'e gyromagnetic factor, and $m_{\ast}$ corresponds to the effective mass of the conduction band electrons in the single-channel one-dimensional semiconducting nanowire. Our goal is to analyze the topological properties of the system under {\color{black}the additional influence of a
time-periodic magnetic field ${\cal B}(t)={\cal B}\cos(\omega t)$, which leads to the time-dependent Zeeman energy term:}
\begin{align}
V_{\rm sm}(t,p_z)=V_{\rm sm}(t+T,p_z)=\frac{\gamma}{2}\,\omega\cos(\omega t)\sigma_3\,,
\end{align}

\noi where we defined the dimensionless driving strength pa\-ra\-me\-ter $\gamma=g\mu_B{\cal B}/\omega$, which quantifies the ratio of the driving amplitude over the driving frequency. {\color{black} It is important to note that in the remainder we consider values for the strengths of the static and time-dependent components of the magnetic field, $B$ and ${\cal B}$, which lead to Zeeman energies in the superconductor that are negligible when compared to the superconducting gap. This is required in order to restrict ourselves to a pa\-ra\-me\-ter regime in which the conventional supercon\-duc\-ting phase remains unaffected. In this manner, the possible scenario of the destruction of the uniform bulk pairing phase and the stabilization of nonuniform Fulde-Ferrell-Larkin-Ovchinnikov~\cite{FF,LO} type of phases can be safely excluded from our investigations.}

{\color{black} Before proceeding, we point out that prior works, such as Ref.~\onlinecite{RoyBasu}, have shown that both MZMs and MPMs are ge\-ne\-ral\-ly accessible for a hybrid system similar to the one considered here, therefore enabling the engineering of effective topological p-wave superconductivity. However, the pro\-xi\-mi\-ty effect and the self-energy of the semiconductor have not been accounted for in previous studies. Hence, crucial aspects concerning the realistic hybrid nanowires have not been addressed. Our formalism fills in this gap and as we demonstrate in the remainder, we do not only recover the emergence of Majorana modes, but we most importantly obtain the topological phase diagrams using the proper self-energy of the semiconductor and evaluate the level broadenings which determine the observability of the theoretically predicted topological phases.}

{\color{black}In fact, as we make plausible in Appendix~\ref{app:AppendixBold} and also cla\-ri\-fy through our upcoming analysis, such a driven hybrid respects a dynamical chiral symmetry~\cite{Yao2017,Kennes,Assili2024} even when the ener\-gy dependence of the self-energy is properly taken into consi\-de\-ra\-tion.} This symmetry implies that the QEO belongs to the BDI symmetry class both with and without the drive. Therefore, both undriven and driven systems support to\-po\-lo\-gi\-cal\-ly nontrivial supercon\-duc\-ting phases with an integer number of MZMs lo\-ca\-li\-zed on each edge of the nanowire. {\color{black}Even more, the driven system simultaneously supports an integer number of MPMs, precisely as occurs when discarding the energy-dependence of the self-energy of the semiconductor~\cite{RoyBasu}. In Sec.~\ref{sec:Invariants}, using the Green function formalism introduced throughout, we provide the construction of the topological invariants that describe the emergence of MZMs and MPMs.}

\subsection{Rotated-Frame Hamiltonian}

In order to improve the convergence of the numerical results, we consider the unitary transformation defined in Eq.~\eqref{eq:UnitarySM} and we transfer the time-dependent Hamiltonian to a new time frame. From Eq.~\eqref{eq:UnitarySMsimple}, we immediately obtain the expression:
\begin{align}
{S}_{\rm sm}(t,p_z;t_0=0)={\rm Exp}\big[-i\gamma\sin(\omega t)\sigma_3/2\big]\,.
\label{eq:UnitarySMexample}
\end{align}

\noi Using the above and Eq.~\eqref{eq:RotatedHSM}, {\color{black}we now end up with} the time-dependent semiconductor Hamiltonian in the new frame:
\bea H_{\rm sm}(t,p_z)&=&\left(\frac{p_z^2}{2m_{\ast}}-\mu\right)\tau_3+E_Z \sigma_3\no\\
&+&\alpha p_z\tau_3\left[ie^{-i\theta(t)}\sigma_--ie^{i\theta(t)}\sigma_+\right],
\label{eq:tildeH_F}
\eea

\noi where we {\color{black}defined} the angle $\theta(t)=\gamma\sin(\omega t)$ along with the raising and lowering spin operators $\sigma_{\pm}=(\sigma_1\pm i\sigma_2)/2$. By employing Eq.~\eqref{eq:FourierSeries}, we expand the above Hamiltonian into a Fourier series with coefficients:
\bea
H_{{\rm sm};n}(p_z)&=&\left[ \left(\frac{p_z^2}{2m_{\ast}}-\mu\right)\tau_3+E_Z \sigma_3\right]\delta_{n,0}\no
\\
&+&\alpha p_z\tau_3 \Big[J_n^+(\gamma)\sigma_2-iJ_n^-(\gamma)\sigma_1\Big],\quad\\\no
\label{eq:SemiHamFloquet}
\eea

\noi where we introduced $J_n^\pm=(J_n\pm J_{-n})/2$. By virtue of the property $J_{-n}=(-1)^nJ_n$, the function $J_n^+$ ($J_n^-$) enters the Hamiltonian only for even (odd) values of $n$.\\

\subsection{Superconducting Proximity Effect and Self-Energy}

We now proceed by inferring the expressions for the self-energy. We can directly read out the result {\color{black}by employing} Eqs.~\eqref{eq:RealSelfieSimple} and~\eqref{eq:BroadeningSimple}, after setting $\beta_{\rm sm}(p_z)=\gamma\sigma_3/2$ and $\zeta=0$. {\color{black}In order to obtain the expression for} the self-energy it is required to evaluate operators of the form $J_n(\gamma\sigma_3/2)$. Based on the pro\-per\-ty of the Bessel functions $J_n(-x)=(-1)^nJ_n(x)$, we can re-write the operator $J_n(\gamma\sigma_3/2)$ in the following fashion:
\begin{align}
J_n(\gamma\sigma_3/2)=\sigma_3^nJ_n(\gamma/2)\equiv e^{-in\pi\frac{1-\sigma_3}{2}}J_n(\gamma/2)\,.
\end{align}

\noi With the help of the above result, we obtain the following expressions for the hermitian and anti-hermitian matrix elements of the retarded self-energy matrix:
\begin{widetext}
\bea
-\widetilde{\Sigma}_{{\rm sm};nm}(\varepsilon,p_z)
&=&\Gamma e^{-i(n-m)\pi\frac{1-\sigma_3}2}\sum_\ell J_{n-\ell}(\gamma/2)\frac{\varepsilon-\ell\omega-\Delta\tau_1}
{\sqrt{\Delta^2-(\varepsilon-\ell\omega)^2}}J_{m-\ell}(\gamma/2)
\,\Theta\Big[\Delta^2-(\varepsilon-\ell\omega)^2\Big]\,,\label{eq:RealSelfieSimpleExample}\\
\Gamma_{{\rm sm};nm}(\varepsilon,p_z)&=&
\Gamma e^{-i(n-m)\pi\frac{1-\sigma_3}{2}}\sum_\ell J_{n-\ell}(\gamma/2)\frac{\varepsilon-\ell\omega-\Delta\tau_1}
{\sqrt{(\varepsilon-\ell\omega)^2-\Delta^2}}J_{m-\ell}(\gamma/2)\,\Theta\Big[(\varepsilon-\ell\omega)^2-\Delta^2\Big]{\rm sgn}(\varepsilon-\ell\omega).\qquad
\label{eq:BroadeningSimpleExample}
\eea
\end{widetext}

\noi From the above expressions, we observe that in the rotated frame the originally spin-independent tunnel-coupling has now acquired a spin structure for the off-dia\-gonal elements with $n\neq m$.

\subsection{Floquet-Green Function Invariants}\label{sec:Invariants}

We are now in a position to explore the topological superconducting phases dictating the hybrid system. Using the de\-fi\-ni\-tion of the QEO in Eq.~\eqref{eq:QuasiEnergyOP}, we define the matrix elements of the QEO for zero and $\pi$ modes as follows:
\begin{align}
{\cal H}_{{\rm sm};nm}^{(s)}(p_z)=\Omega_{nm}^{(s)}+H_{{\rm sm};n-m}(p_z)+\widetilde{\Sigma}_{{\rm sm};nm}^{(s)}(p_z)\,,
\end{align}

\noi where $s=\{0,\pi\}$ and we introduced:
\bea
\Omega_{nm}^{(s)}&=&\big(n-s/2\pi\big)\omega\delta_{n,m}\,,\\
\widetilde{\Sigma}_{{\rm sm};nm}^{(s)}(p_z)&=&\widetilde{\Sigma}_{{\rm sm};nm}(\varepsilon=s\omega/2\pi,p_z)\,.
\eea

The QEO defines a matrix $\hat{{\cal H}}_{\rm sm}^{(s)}(p_z)$ in the extended Floquet-Hilbert space $\mathbb{S}\equiv\mathbb{F}\bigotimes\mathbb{H}$, i.e., the so-called Sambe space~\cite{Sambe}, where $\mathbb{H}$ represents the state space of a quantum system and  $\mathbb{F}$ is a space spanned by all square-integrable $T$-periodic time-dependent functions. While the quasi-energy matrix has infinite dimensions, in practice the desired band structure and topological invariants are generally obtained up to a finite truncation. As we made plausible in Appendix~\ref{app:AppendixBold}, the QEO features a dynamical chiral symmetry and belongs to class BDI, which allows for two winding numbers, i.e., $\nu_s\in\mathbb{Z}$, with $s=0$ and $s=\pi$ that predict the number of edge MZMs and MPMs, respectively. In our prior work~\cite{Assili2024}, we discussed such symmetry scenarios in detail and identified the ge\-ne\-ral transformation which renders the dynamical chiral symmetry operator block dia\-go\-nal and, simultaneously, the QEO block-off dia\-go\-nal, so that:
\begin{align}
\underline{\cal{H}}_{\rm sm}^{(s)}(p_z)=
\begin{pmatrix}
\hat{0}&\hat{A}_s(p_z)\\
\hat{A}_s^\dag(p_z)&\hat{0}\end{pmatrix}
\phd{\rm with}\phd
\underline{\Pi}_s=
\begin{pmatrix}
\hat{\mathds{1}}&\hat{0}\\
\hat{0}&-\hat{\mathds{1}}\end{pmatrix}.\label{eq:BlockOffDiag}
\end{align}

\noi The corresponding topological invariants $\nu_s$  are given as:
\begin{align}
\nu_s=\frac{i}{2\pi}\int_{-\infty}^{+\infty}dp_z\ph{\rm tr}\left[\hat{A}_s^{-1}(p_z)\frac{d \hat{A}_s(p_z)}{dp_z}\right],
\label{eq:windingnumer}
\end{align}

\noi where ``tr" defines the matrix trace in the space of $\hat{A}_s(p_z)$. In one-dimensional systems, one can alternatively express the topological invariants as the winding number of the determinant of $\hat{A}_s(p_z)$. Therefore, if we define $\det[\hat{A}_s(p_z)]=\big|\det[\hat{A}_s(p_z)]\big|e^{-i\varphi_s(p_z)}$, we can express the topological inva\-riants as $\nu_s=\int_0^{2\pi}\, d\varphi_s/2\pi$. Finally, we remind the reader of the fact that $\nu_\pi$ is zero in the high-frequency driving regime.

\subsection{Analysis for Low Driving Frequencies}

Having set our Floquet framework, we now proceed and examine the topological phase diagrams by eva\-lua\-ting the respective invariants. We first consider the case in which $\omega\ll\Delta$.
As it is customary, we truncate the Floquet space and consider only a few {\color{black}Floquet zones centered at the quasi-energy of inte\-rest~\cite{YangZhesen} that span an energy regime smaller than $2\Delta$. The criterion for the optimal truncation in our numerical calculations is obtained by identifying when convergence is achieved. That is, by inferring beyond which number of Floquet modes included our results do not any longer change, within the nu\-me\-ri\-cal accuracy chosen. It is worth noting that the convergence of the QEO method in frequency space typically requires a small number of Floquet modes thanks to the localization of the Floquet eigenstates in frequency space, which is analogous to the Wannier-Stark lo\-ca\-li\-za\-tion taking place in real space.}

In the truncated space, the retarded and advanced self-energies are purely hermitian, since there is no broa\-de\-ning. In the following, we first discuss the quasi-energy spectra and subsequently the respective topological phase diagrams.

\begin{figure}[t!]
\begin{center}
\includegraphics[width=0.494\columnwidth]{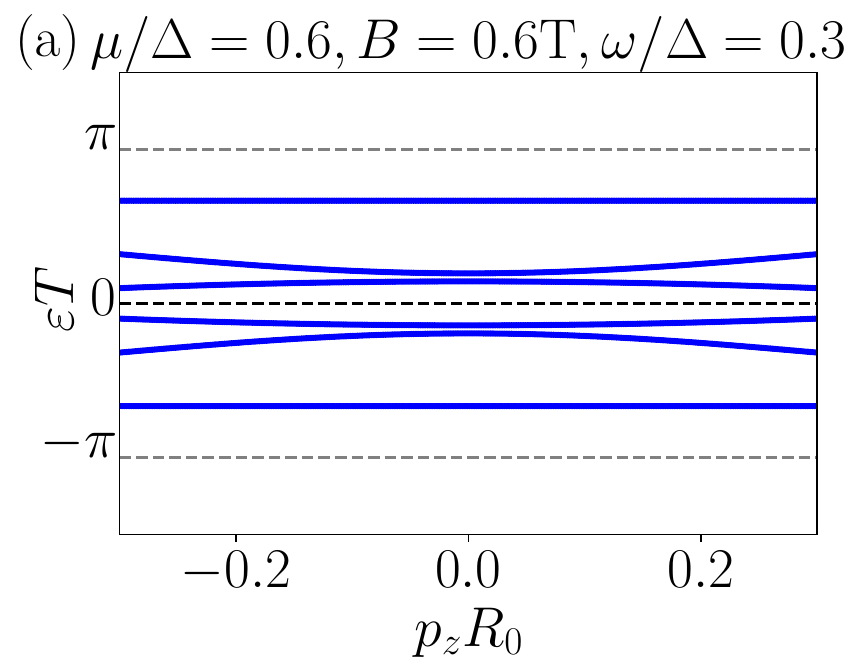}
\includegraphics[width=0.494\columnwidth]{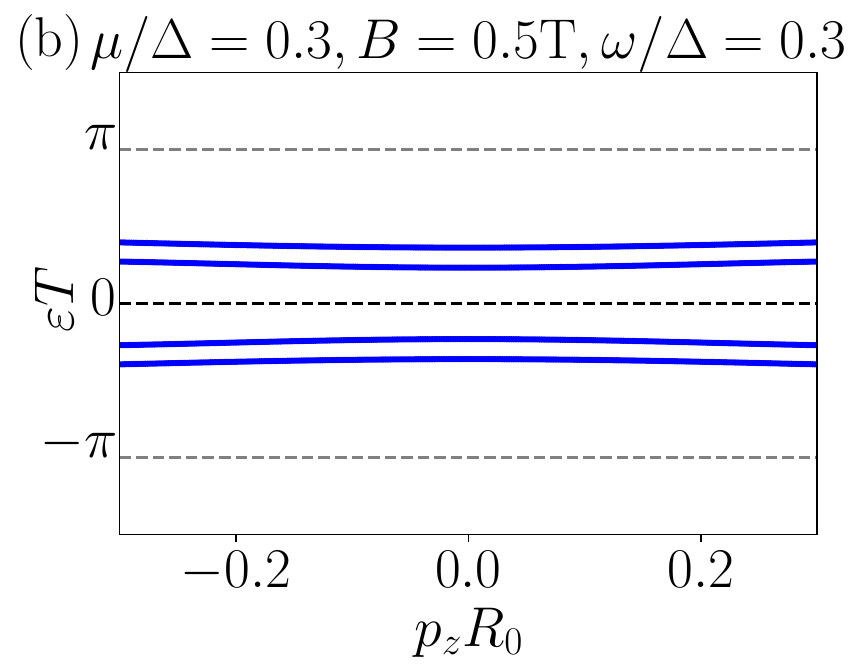}
\includegraphics[width=0.494\columnwidth]{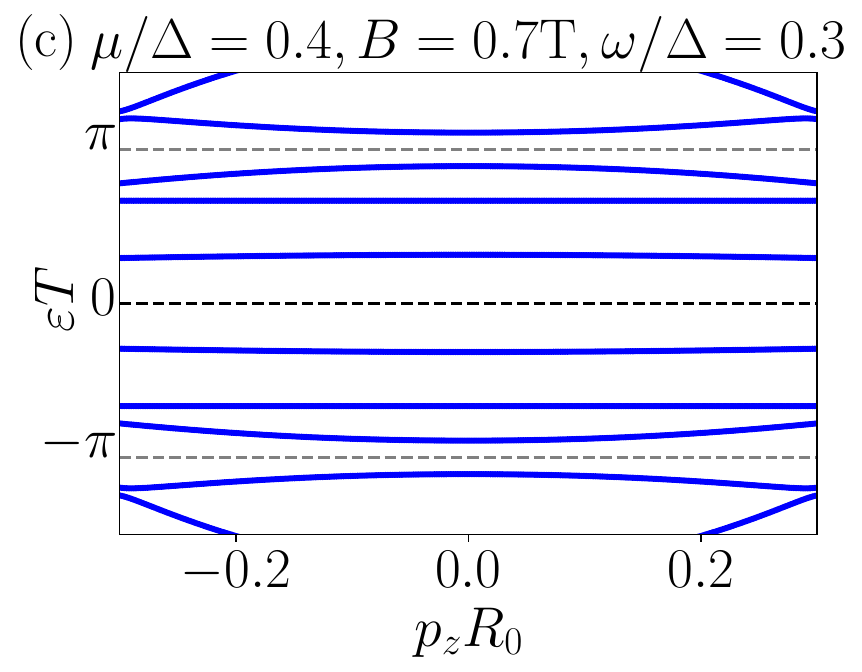}
\includegraphics[width=0.494\columnwidth]{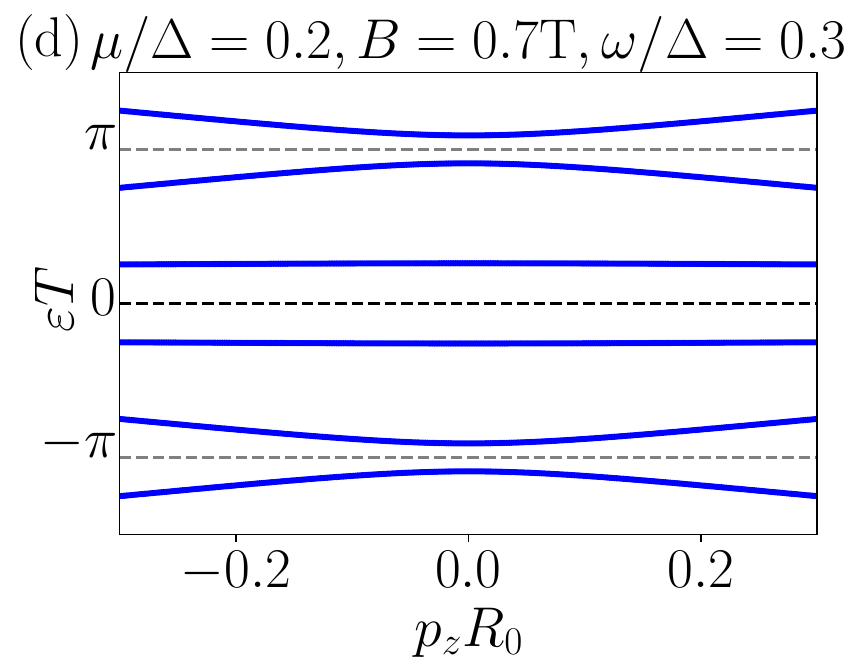}
\end{center}
\caption{Quasi-energy spectra obtained numerically for the following values of che\-mi\-cal potential and magnetic field, i.e., $(\mu/\Delta, B ({\rm T}))=\{(0.6,0.6),(0.3,0.5),(0.4,0.7),(0.2,0.7)\}$ and driving frequency $\omega/\Delta=0.3$. While in the absence of the drive, the superconductor-induced self-energy on the semiconductor solely renormalizes the band structure of individual Floquet modes, in the presence of the drive, the self-energy also acquires nonzero off-diagonal elements in Floquet space, thus, further contributing to the hybridization of the Floquet bands. Note that the gaps around zero and $\pi$ quasi-energies are protected by the presence of a chiral symmetry which is preserved even when the hybrid is driven.}
\label{fig:Figure3}
\end{figure}

\subsubsection{Quasi-Energy Spectra}

For all the numerical calculations throughout this work, we use dimensionless forms of the Hamiltonian and self-energy in Eqs.~\eqref{eq:SemiHamFloquet},~\eqref{eq:RealSelfieSimpleExample}, and~\eqref{eq:BroadeningSimpleExample}, where all ener\-gy scales are expressed in units of the pairing gap of the bulk superconductor $\Delta$. After carrying out this re\-scaling, it is also convenient to introduce the charac\-te\-ri\-stic Rashba SOC strength $\alpha_0=\sqrt{\Delta/2m_{\ast}}$ and the length scale $R_0=1/\sqrt{2m_{\ast}\Delta}$, along the lines of Ref.~\onlinecite{Sole}. For $\Delta= 400\,\mu$eV, $\alpha=0.1\,{\rm eV}$\AA\,  and $g=15$, we find $\alpha/\alpha_0=0.38$, $R_0=64.5$ nm, $\alpha_0=0.26\,{\rm eV}$\AA, and $E_Z=g\mu_B B/2=1.08B$, with the Zeeman field expressed in teslas. Lastly, the tunnel-coupling parameter $\Gamma$ and the dimensionless drive amplitude $g\mu_B{\cal B}/\Delta$ are set to the values 1 and 0.3 respectively in all numerical calculations that follow. We now proceed and derive the quasi-energy spectrum of the system by numerically computing the roots of the QEO $\hat{{\cal H}}_{\rm sm}^{(s)}(\varepsilon,p_z)$ which is inferred from Eq.~\eqref{eq:QuasiEnergyOP}. For this purpose we numerically evaluate the zeros of the determinant of the matrix given by the QEO. {\color{black}In our numerical calculations, we set the truncation of the QEO to 7$\times 7$ in the Floquet space, resulting in a matrix with the dimension 28 $\times$ 28 in Sambe space. By evaluating the determinant of the matrix, one can determine the eigenvalues at which the determinant vanishes. It is worth noting that the self-energy depends on the quasienergy $\varepsilon$ which implies that the obtained results are sensitive to the pairing gap $\Delta$ and the drive frequency $\omega$.
}

Figure~\ref{fig:Figure3} depicts the numerically-obtained quasi-energy spectra in the various values of chemical potential and magnetic field, when con\-si\-de\-ring a low driving frequency $\omega/\Delta=0.3$. We observe the opening of energy gaps at both zero and $\pi$ quasi-energy regions which are prerequisite {\color{black}for obtaining} MZMs and MPMs. As illustrated in the panels of Fig.~\ref{fig:Figure3}, the proximity-induced self-energy leads to the substantial reconstruction of the quasi-energy band structure. These effects are extremely sensitive to variations of the chemical potential and magnetic field. Nonetheless, we emphasize that in spite of the strong impact of the self-energy, the emergence of gaps around zero and $\pi$ quasi-energies are protected by the dy\-na\-mi\-cal chiral symmetry of the system~\cite{Yao2017,Kennes,Assili2024}.

\begin{figure}[t!]
\begin{center}
\includegraphics[width=0.494\columnwidth]{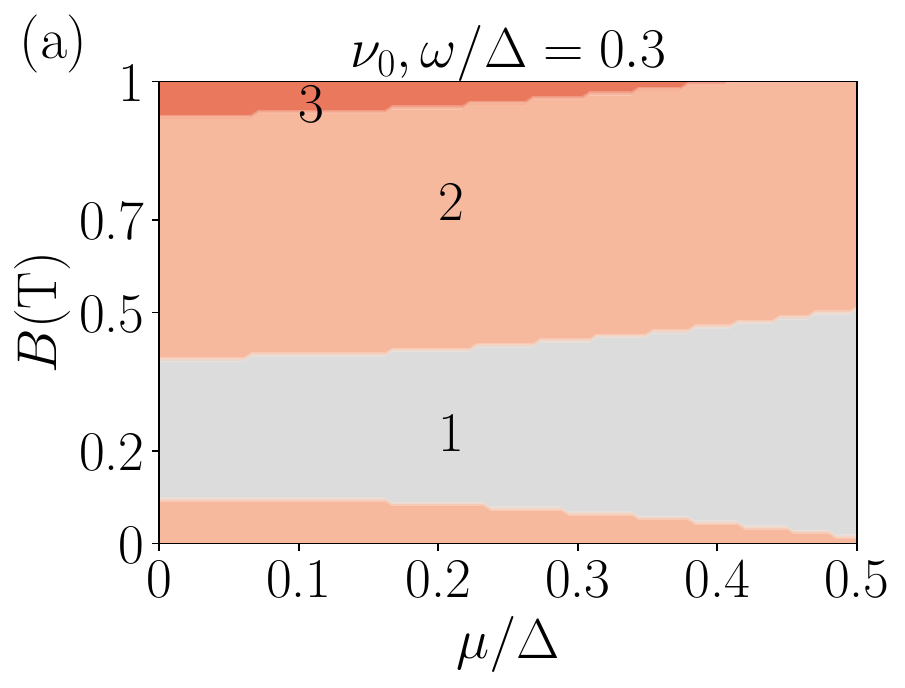}
\includegraphics[width=0.494\columnwidth]{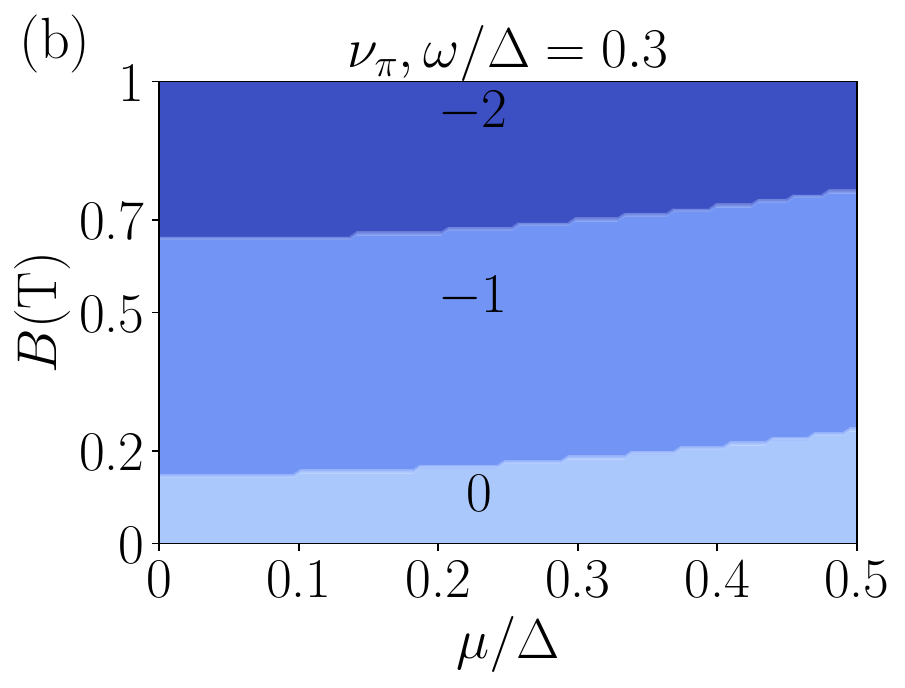}\\
\end{center}
\caption{Numerically-evaluated topological phase diagrams using the self-energy formalism presented in this work. (a) and (b) depict the winding numbers correspon\-ding to MZMs and MPMs, respectively, for a driving frequency $\omega/\Delta=0.3$. The numerical results show that both MZMs and MPMs exist in the low driving frequency regime. In addition, we predict phases with multiple Majorana edge modes, {\color{black}which are otherwise inaccessible in the undriven hybrid}.}
\label{fig:Figure4}
\end{figure}

\subsubsection{Topological Phase Diagrams}

In this section, we employ our formalism to analyze the effects of the harmonic drive on the topological pro\-per\-ties of the hybrid system. We numerically evaluate the win\-ding numbers $\nu_{0,\pi}$ for MZMs and MPMs, which are both in principle nonzero in the low-frequency regime. {\color{black} We note that the number of matrix elements of the QEO in the Sambe space is, in principle, infinite, and we therefore truncate it up to a limit where convergence is achieved. Importantly, expressing the quasienergy operator in a rotated time frame improves convergence, allowing us to truncate at an optimal limit. Moreover, the truncations corresponding to the zero and $\pi$ modes are odd and even, respectively, and preserve chiral symmetry in the Sambe space. The details of this numerical procedure have been discussed in our previous study~\cite{Assili2024}}. In our numerical calculations we truncate the QEO and retain the $0,\pm 1,\pm 2$ Floquet modes. {\color{black}This process results in a $20\times 20$ and $16\times 16$ matrices associated with zero and $\pi$ modes, respectively, in Sambe space}. Furthermore, we also truncate the sum over $\ell$ appearing in Eq.~\eqref{eq:RealSelfieSimpleExample} to the subspace of $0,\pm 1,\pm 2,\pm 3 $ Floquet modes. Given these assumptions, we find that the self-energy operator is fully hermitian in this regime. In fact, it is in this low-driving-frequency regime that the proximity-induced contribution to the quasi-energy spectrum of the semiconductor becomes more pronounced.

In Fig.~\ref{fig:Figure4}, we show the topological phase diagrams for MZMs and MPMs as functions of  chemical potential and Zeeman field. For these numerical calculations we set $\omega/\Delta=0.3$, which ensures that the self-energy remains fully hermitian for Floquet harmonics up to $\ell=\pm2$. These terms contain the dominant contributions, as we consider a small drive amplitude, for which the Bessel functions entering the expressions of the self-energy decay rapidly as the order of the Bessel function increases. Therefore, in the weak driving amplitudes and low dri\-ving frequency regime considered here, the self-energy operator is essentially hermitian, thus, here the broa\-de\-ning effects are negligible, in accordance with Ref.~\onlinecite{YangZhesen}.

\begin{figure}[t!]
\begin{center}
\includegraphics[width=0.494\columnwidth]{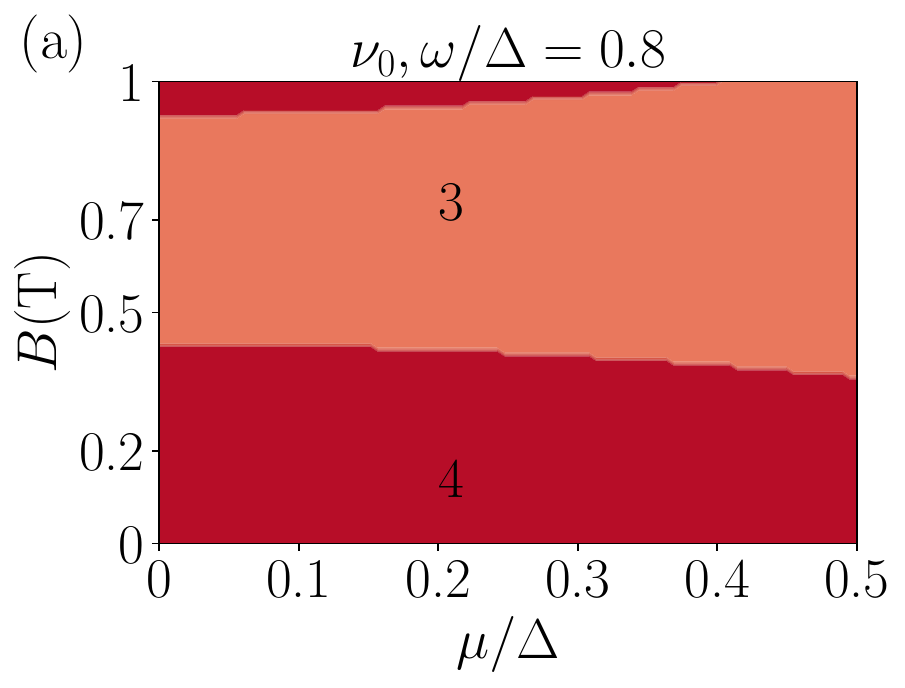}
\includegraphics[width=0.494\columnwidth]{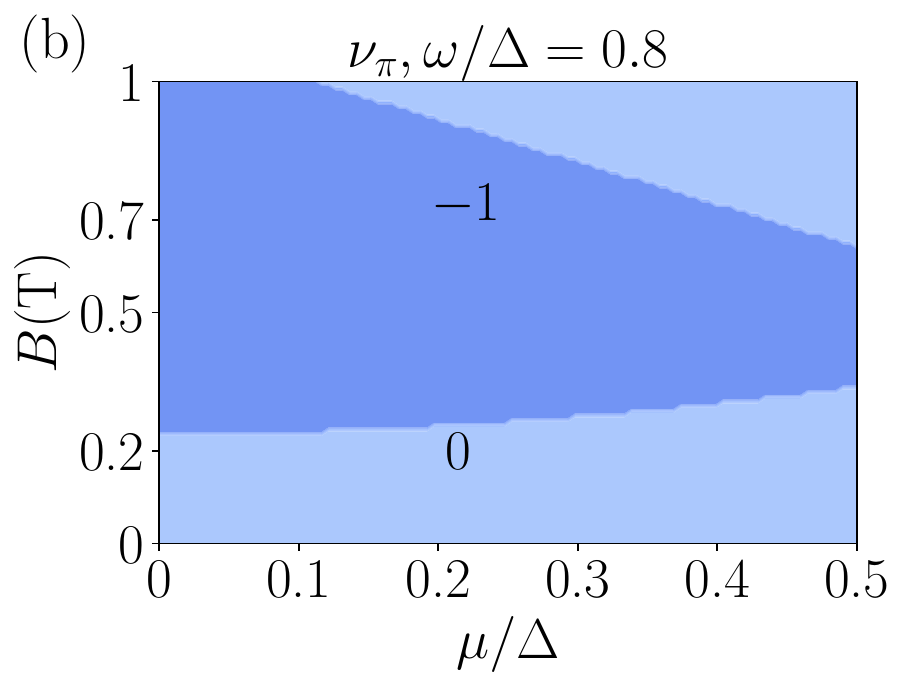}
\end{center}
\caption{(a) and (b) present our numerical calculations of the topological phase diagrams for MZMs and MPMs, as functions of the chemical potential and the Zeeman field for the intermediate driving frequency $\omega/\Delta=0.8$. Quite remar\-kably, in (a) we observe the emergence of topologically non-trivial phases harboring multiple MZMs per edge already in the range of zero and generally low Zeeman field values. Such a trend was also observed for lower frequencies. See Fig.~\ref{fig:Figure3}(a). In contrast, a threshold Zeeman field is required for MPMs in both low and intermediate values of driving frequency.}
\label{fig:Figure5}
\end{figure}
As it is also depicted in Fig.~\ref{fig:Figure4}, we observe that both Majorana sectors undergo topological transitions at lower values of the Zeeman field, when compared to the si\-tua\-tion observed in the undriven case, where the topological transition for $\varepsilon=0$ occurs at $E_Z=\Delta$ when the chemical potential is set to zero. {\color{black}Moreover}, we note that both  topological phase diagrams contain extensive regions with winding numbers higher than $\pm1$. {\color{black}Remarkably, the emergence of multiple Majorana modes at both zero and $\pi$ quasi-energies is a feature shared with pre\-vious related works~\cite{RoyBasu}, even though in those the superconduc\-ting pro\-xi\-mi\-ty effect was solely accounted for via the introduction of a pairing gap in the semiconductor.}

\subsection{Analysis for Intermediate Driving Frequencies}

We now consider the case in which the driving frequency is set to $\omega/\Delta=0.8$. Here, it is expected that the broadening effect is significant compared to the low frequency regime. The panels (a) and (b) in Fig.~\ref{fig:Figure5} show the resulting to\-po\-lo\-gi\-cal phase diagrams associated with MZMs and MPMs. In this frequency range, the driven hybrid harbors non-trivial edge modes for a wide range of chemical potentials and Zeeman fields. Moreover, the topological phase diagram associated with MZMs reveals a high winding number phase emerging at low Zeeman fields. In contrast, the MPMs phase diagram does not support higher numbers in this parameter window.

\begin{figure}[t!]
\begin{center}
\includegraphics[width=0.494\columnwidth]{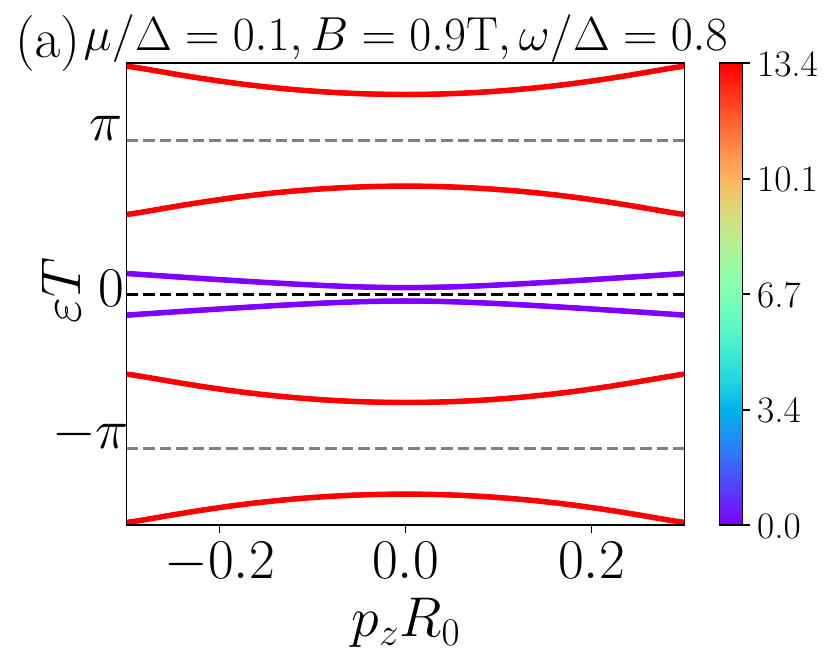}
\includegraphics[width=0.494\columnwidth]{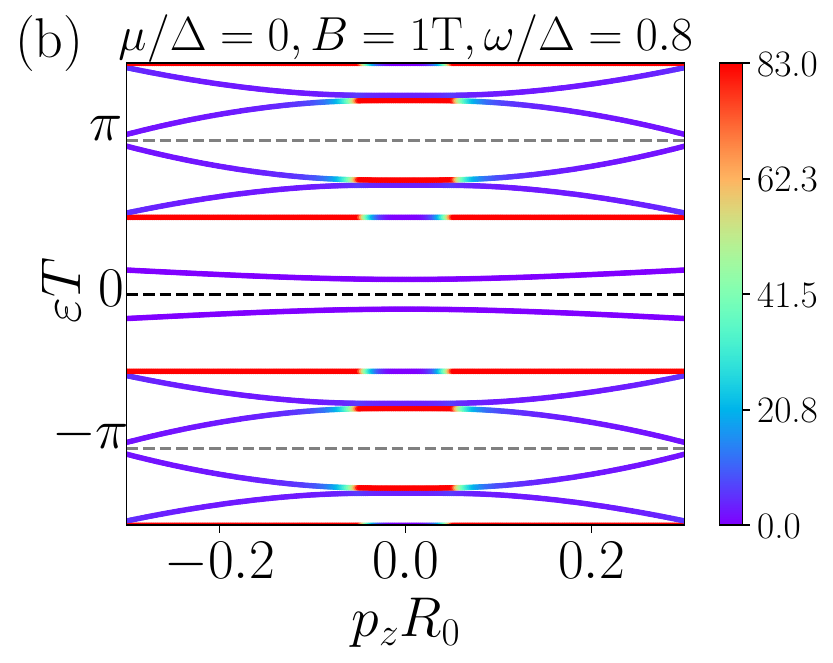}
\includegraphics[width=0.494\columnwidth]{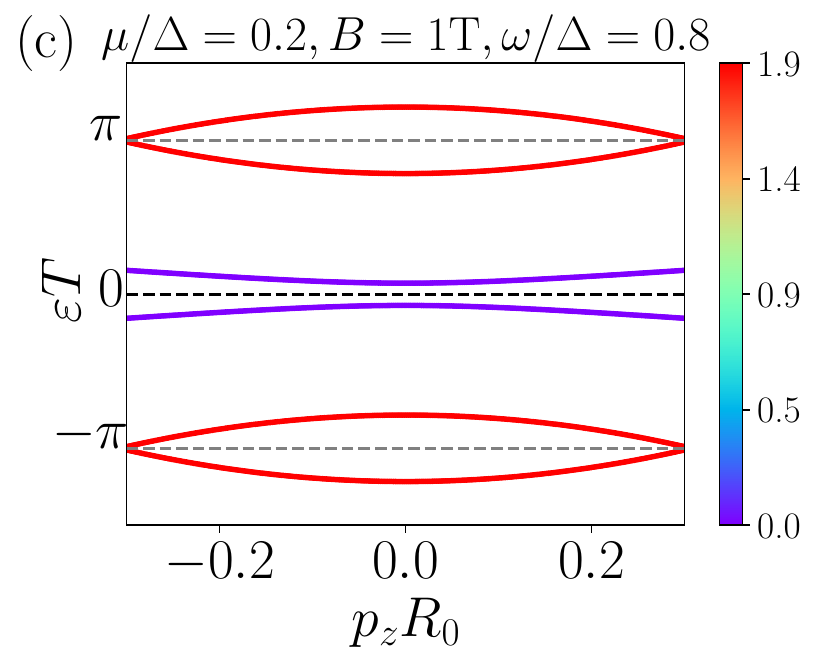}
\includegraphics[width=0.494\columnwidth]{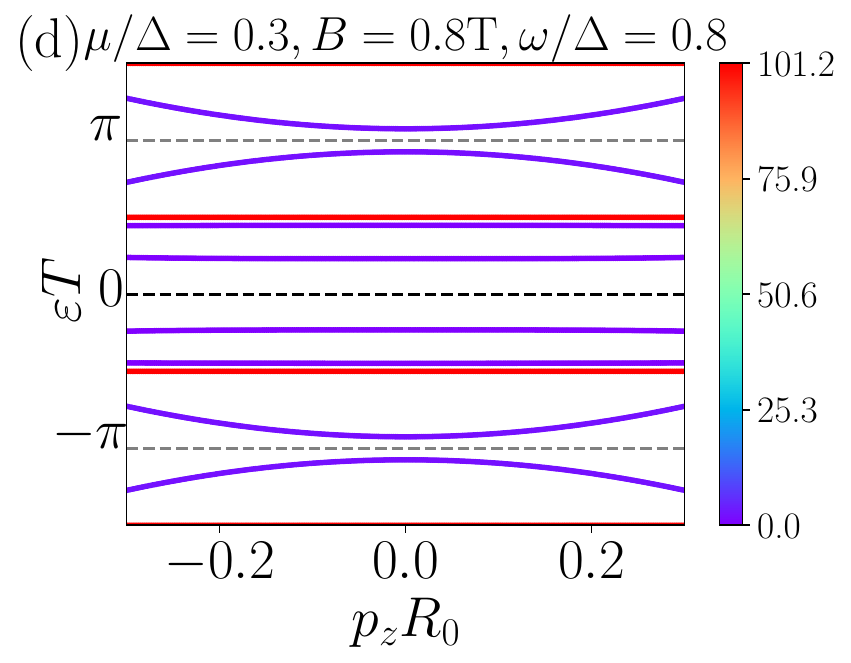}
\end{center}
\caption{Numerical evaluation of the quasi-energy band structure, color-coded with the corresponding value of the associated level broadening for $\omega/\Delta=0.8$. Our results show that the quasi-energy bands correspon\-ding to MPMs are significantly more sensitive to the broa\-dening compared to the MZMs for this choice of driving frequency. Therefore, as it was also pointed out in Ref.~\onlinecite{YangZhesen}, the to\-po\-lo\-gi\-cal properties of the $\pi$-modes are difficult to observe in experiments due to the substantial level broadening.}
\label{fig:Figure6}
\end{figure}

We now treat the effects of the broadening induced by the anti-hermitian part of the self-energy on the quasi-energy bands. For this purpose, we numerically calculate the projection of the broadening matrix given in Eq.~\eqref{eq:BroadeningSimple} onto the eigenstates of the QEO. Our numerical results are shown in Fig.~\ref{fig:Figure6} and are plotted for several va\-lues of the chemical potential and the Zeeman field with $\omega/\Delta=0.8$. We observe that for these choices of pa\-ra\-me\-ters, the quasi-energy bands near zero ener\-gy are {\color{black} very weakly affected by the broa\-de\-ning. However, for high ener\-gies and specifically around $T\varepsilon=\pi$, the broa\-de\-ning tends to become increasingly important. As a result, to\-po\-lo\-gi\-cal MPM phases appear to be more challenging to observe in experiments. This is probably because in this frequency regime the bands responsible for the MPMs lie closer to the pairing gap edge.} {\color{black}The disparity between the strengths of the broadening for zero and $\pi$ quasi-energies is expected judging from Eq.~\eqref{eq:BroadeningSimple}. From there, one finds that the broa\-de\-ning is a rather involved function of the quasi-energy, the driving frequency and amplitude, and the pairing gap. Hence, no strict universal trend is expected to emerge. For instance, one observes that in Fig.~\ref{fig:Figure6}(d) the broadening near $\pi$ quasi-energy is much smaller than the ones obtained for the cases (a)-(c).}

\subsection{Analysis for High Driving Frequencies}

We now move on and consider that the frequency is sufficiently high so that each $\ell\neq0$ Floquet mode belonging to the truncated Floquet space satisfies $|\varepsilon-\ell\omega|\gg\Delta$.

\subsubsection{Approximate Analytical Results}

Since, in this driving regime only MZMs become accessible, we first employ Eq.~\eqref{eq:HighFreqFreenRet} to obtain the approximate expression for the inverse retarded matrix Green function for the zero-th Floquet mode. While a detailed eva\-lua\-tion is presented in Appendix~\ref{app:AppendixCold}, here we summarize the main effects.

{\color{black}It is well established that the topological phase transitions of the undriven system appear only due to the gap closing at zero quasi-energy and at zero momentum $p_z=0$. Since the Rashba SOC term va\-ni\-shes at $p_z=0$, this implies that the strength of the Rashba term does not influence the topological phase diagram. This result na\-tu\-ral\-ly extends to the driven case due to the specific type of Zeeman-field drive considered here.} {\color{black}This is because, when it comes to the hermitian part of the self-energy, this only renormalizes the strength of the Rashba SOC, which now becomes $\alpha\mapsto\tilde{\alpha}=\alpha J_0(\gamma)$.} {\color{black}Moreover, other possible topological phase transitions that could arise in the driven system due to the vanishing of the Bessel function $J_0(\gamma)$ and yield a sign change in the topological inva\-riant, are not relevant here since we have restricted ourselves to a weak drive amplitude.} {\color{black}Hence, we expect for the to\-po\-lo\-gi\-cal phase diagram in this regime to exhibit negligible dif\-fe\-ren\-ces compared to the one obtained for the undriven hybrid.}

We now examine the anti-hermitian contribution of the self-energy. From Appendix~\ref{app:AppendixCold}, we find that in the weak drive-amplitude $\gamma\ll1$ and low-energy $|\varepsilon|\ll\Delta$ limits,  the broa\-de\-ning obtains the energy-independent form:
\begin{align}
\Gamma_0(p_z)\approx \Gamma\frac{\gamma^2}{2}\left[\left(\frac{\Gamma/2}{\omega}\right)^2+\left(\frac{\alpha p_z}{\omega}\right)^2\right].
\label{eq:Broad}
\end{align}

\noi From the above, we conclude that for small momenta $p_z$ the broadening is constant and is determined by the first contribution, which originates from Cooper-pair-mediated transitions. Instead, for large momenta the broadening is dominated by the second term which is sourced from the Rashba SOC.

\subsubsection{Numerical Results}

\begin{figure}[t!]
\begin{center}
\includegraphics[width=0.72\columnwidth]{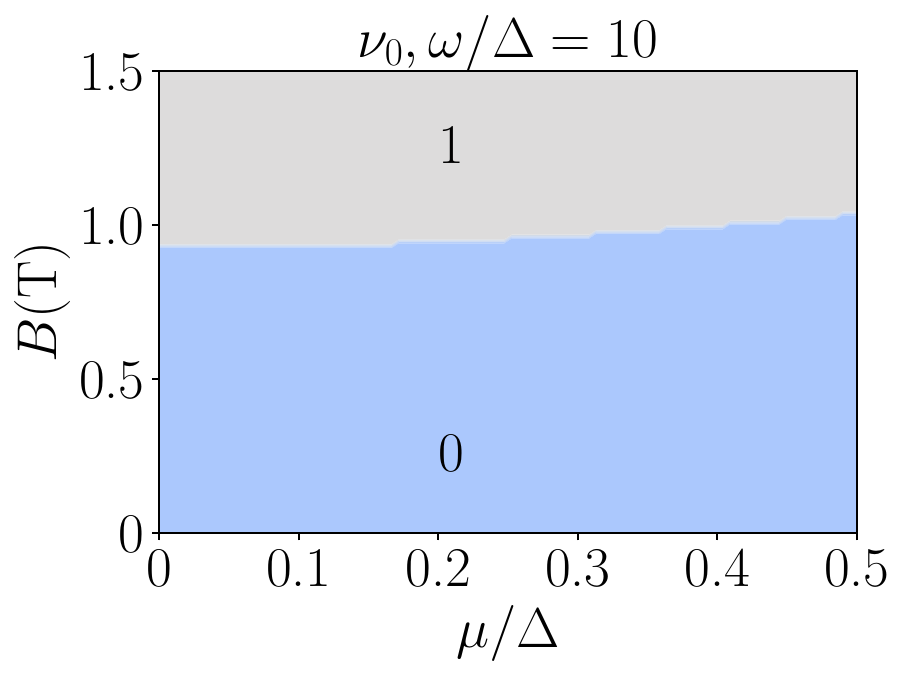}
\end{center}
\caption{Numerical calculation of the topological phase dia\-gram in the high frequency regime. The numerical results show the winding number correspon\-ding to MZMs as a function of chemical potential and magnetic field. The obtained topological phase diagram in the high frequency regime is essentially identical to the one found in the undriven system.}
\label{fig:Figure7}
\end{figure}

In Fig.~\ref{fig:Figure7}, we present the outcomes of our exact nu\-me\-ri\-cal calculations obtained in the high-frequency regime with $\omega/\Delta=10$. Notably, we observe that the topological phase diagram is analogous to the one obtained for undriven case, which is also in accordance with our ana\-ly\-ti\-cal conclusions of the above paragraphs. Furthermore, from our analytics we also find that the resulting broa\-dening is very small, which is consistent with the predictions of Eq.~\eqref{eq:Broad}{\color{black}, since $\gamma/\omega\ll1$ and we are interested in momenta $p_z\sim0$}. Concluding our discussion, it is important to remark that since in the high-frequency regime {\color{black}the system behaves in a quasi-static manner}, one can demonstrate that the topological phase diagram is unaffected by the drive by formulating a topological invariant using a causal Euclidean Green function which also incorporates both effects of Rashba SOC renormalization and broa\-de\-ning. This approach is briefly discussed in Appendix~\ref{app:AppendixDold}  {\color{black} and transparently shows that the broadening neither affects the outcome of the topological invariants nor the emergence of phase transitions.}

\section{Discussion and Outlook}

We bring forward a general theoretical framework tailored for the description of the Floquet topological phases of driven superconductor-semiconductor hybrids. Our formulation relies on the Green function technique and the analytical eva\-lua\-tion of the self-energy of the semiconductor due to the super\-con\-duc\-ting proximity effect in the additional presence of the drive. Our approach provides a comprehensive description of the quasi-energy renormalization, topological inv\-a\-riants, and broadening effects induced by the superconducting bath on the semiconductor. Specifically, a key aspect of our work is that we prescribe how to define the quasi-energy operator by incorporating the hermitian part of the self-energy of the semiconductor, while we propose to infer the respective level broadenings by projecting the anti-hermitian part of the self-energy onto the quasi-energy eigenvectors. {\color{black}Our general method and results apply to a wide range of superconductor-semiconductor hybrids and types of dri\-vings, therefore providing a fundamental basis for future studies on driven topological superconductivity and Floquet invariants in the pre\-sen\-ce of self-energy effects.}

{\color{black}We remark that in order to define and evaluate the Floquet topological invariants it is not necessary to derive the Keldysh part of the Green function. Indeed, the topological invariants are solely associated with the spectral properties of the sy\-stem, which are fully encoded in the retarded and advanced Green functions. These further encompass information regarding the level broadenings of the quasi-energy dispersions, which provide a measure for the observability of the to\-po\-lo\-gi\-cal phase diagram inferred using the topological invariants.

Investigating the behavior of the Keldysh component of the Green function would be necessary when discussing physical observables, such as, the particle density or the density of states. The study of such physical quantities for these hybrids has been previously carried out in Refs.~\onlinecite{LiuLevchenkoLutchyn,YangZhesen,Forcellini}. However, studying these physical quantities does not allow to disen\-tan\-gle the topological invariants from the consequences of broa\-de\-ning. This is exactly what our approach achieves. Here, one can have a clear picture of the topology of the bands along with the effects of broadening in a separate fashion, therefore offering the opportunity to reach a deeper insight concerning the topological properties. Having as our key priority to address the construction of  topological invariants of this general class of driven systems, a number of interesting questions were unavoidably left outside of the scope of our analysis, such as,  heating and other nonequilibrium aspects which are typically considered for periodically driven systems~\cite{LiuLevchenkoLutchyn, Weidinger,Lubatsch,Frank,Mosallanejad}. Such important aspects are complementary to the approach introduced in this work and are left for a future study.
}

{\color{black}In this manuscript, we additionally exemplify our approach by applying it to a hybrid system which consists of a Rashba nanowire which is tunnel-coupled to a bulk conventional superconductor and is under the simultaneous influence of a combination of static and time-periodic components for a Zeeman field oriented along its axis.} {\color{black}In the undriven Rashba nanowire system, the topological transition leading to zero Majorana modes is controlled by three key parameters: the pairing gap, the chemical potential, and the static Zeeman ener\-gy. For this reason, we decided to consider an additional time-periodic Zeeman-field component, as this is expected to drastically affect the topological properties of the system. To investigate this model, we employed a new time frame, where we solve an equivalent but better-convergent problem, since the time-periodic dependence is now transferred to the Rashba term. Hence, this mapping allows us to also address the case in which the driving is a result of a time-dependent electric field. In this case, the time-periodic drive can be view as the result of an ac Stark effect, thus establishing a link to the ac Zeeman effect considered in the original time frame.}

{\color{black}Our combined approximate-analytical and numerical approaches demonstrate that, in} the low-frequency regime, the self-energy is fully hermitian and solely leads to quasi-energy renormalization effects. In this regime, both Majorana zero modes and Majorana $\pi$ modes emerge over a wide range of chemical potentials and Zeeman fields, while they are unaffected by level broa\-de\-ning, since the anti-hermitian part of the retarded self-energy is negligible. In the intermediate-frequency regime, however,  broa\-de\-ning effects are present. In fact, our numerics show that these are generally substantial and suppress the observability of topological phases associated with Majorana $\pi$ modes. {\color{black}Hence, in order to engineer robust Majorana $\pi$ modes, the desired strategy is to drive the system with low frequencies where broadening is generally negligible.}
{\color{black}On the other hand,} Majorana zero modes remain robust and immune to broa\-de\-ning across all frequency ranges.

{\color{black}The low- and intermediate-frequency regimes appear the most prominent to observe phenomena without static analogs. For instance, in the presence of the driving field one expects the emergence of ac Majorana-Josephson effects mediated by Majorana $\pi$ modes~\cite{Kundu}, thus extending the ones that have been discussed for the undriven case~\cite{Kitaev}. Even more, in the case of Zeeman-field-drive considered here, novel types of magnetically-resolved Majorana-Josephson effects and applications in super-spintronics can also become accessible. These are worthy of future exploration, since they can extend in a nontrivial fashion phenomena that have been previously discussed for the respective undriven nanowire hybrids~\cite{KotetesJosephson,JiangJosephson,PientkaJosephson,MercaldoJosephson}.}

In the high-frequency regime, the system effectively mi\-mics its undriven counterpart since Majorana $\pi$ modes no longer survive, and the obtained phase diagram for  Majorana zero modes becomes essentially identical to the one dictating the hybrid in the absence of the drive. {\color{black}The reason behind this behavior is that, in the high-frequency regime, the effect of the time-periodic Zeeman field in the new time frame is to solely renormalize the strength of the Rashba term, which is known already from the analysis of the undriven Rashba nanowire model not to affect the to\-po\-lo\-gi\-cal phase transition~\cite{LutchynPRL,OregPRL}, but only to determine the spatial profile of the arising Majorana modes. By exploring the high-frequency driving regime using the Euclidean Green function formalism, we further explicitly prove that the broadening does not affect the topology of the hybrid.}

\begin{acknowledgments}
We wish to thank H.-Q. Xu for valuable discussions and for his strong support while carrying out this project.
\end{acknowledgments}

\section*{AUTHOR DECLARATIONS}

\section*{C\lowercase{onflict of} I\lowercase{nterest}}
The authors have no conflicts to disclose.

\section*{A\lowercase{uthor} C\lowercase{ontributions}}
The two authors contributed equally to this work.\\

\bt{Mohamed Assili:} Conceptualization (equal); Data curation (equal); Formal
analysis (equal); Funding acquisition (equal); Investigation (equal);
Methodology (equal); Project administration (equal); Resources
(equal); Software (equal); Supervision (equal); Validation (equal);
Visualization (equal); Writing – original draft (equal); Writing –
review \& editing (equal). \bt{Panagiotis Kotetes:} Conceptualization (equal);
Data curation (equal); Formal analysis (equal); Funding acquisition
(equal); Investigation (equal); Methodology (equal); Project administration (equal); Resources (equal); Software (equal); Supervision
(equal); Validation (equal); Visualization (equal); Writing – original
draft (equal); Writing – review \& editing (equal).

\section*{Data Availability}

The data that support the findings of this study are available within the article.

\appendix

\section{Self-Energy Calculation}\label{app:AppendixA}

The assumptions considered in Sec.~\ref{sec:Typical} allow us to simplify the expression in Eq.~\eqref{eq:Self-Energy} in the following fashion:
\bea
&&\Sigma_{{\rm sm};nm}(\varepsilon,\bm{p})={\rm T}^2\,e^{i\zeta(n-m)} \sum_\ell J_{n-\ell}[\beta_{\rm sm}(\bm{p})]\,\,\quad\no\\
&&\,\,\,\quad\times\tau_3\left[\int d\bm{q}\,G_{{\rm sc};\ell}(\varepsilon,\bm{p},\bm{q})\right]\tau_3\,J_{m-\ell}^\dag[\beta_{\rm sm}(\bm{p})],
\label{eq:Self-Energy_Simple}
\eea

\noi where we have already made use of the fact that here $V_{\rm sc}(t,\bm{p})=0$, which in turn implies that the Green function of the superconductor is diagonal in Floquet space. Based on these grounds we also introduced the shorthand notation $G_{{\rm sc};\ell s}(\varepsilon,\bm{p},\bm{q})=G_{{\rm sc};\ell}(\varepsilon,\bm{p},\bm{q})\delta_{\ell,s}$.

We now proceed with the evaluation of the momentum integral appearing in Eq.~\eqref{eq:Self-Energy_Simple}. As we  already mentioned earlier, in these hybrids we have the hierarchy $|\bm{p}|^2/2m_0\ll E_F$, which allows us to drop the $\bm{p}$ dependence in the Floquet-Green function of the superconductor and find:
\begin{align}
\int d\bm{q}\,G^{R,A}_{{\rm sc};\ell}(\varepsilon,\bm{p},\bm{q})\simeq\int d\bm{q}\,\frac{\mathds{1}}{\varepsilon-\ell\omega\pm i0^+-\xi(\bm{q})\tau_3-\Delta\tau_1}\no\\
=-\pi\nu_F\int_{-E_F}^{+\infty}\frac{d\xi}{\pi}\,\frac{\varepsilon-\ell\omega+\xi\tau_3+\Delta\tau_1}{\xi^2+\Delta^2-(\varepsilon-\ell\omega)^2\mp i{\rm sgn}(\varepsilon-\ell\omega)0^+},
\end{align}

\noi where by using standard manipulations we converted the $\bm{q}$-integral to a $\xi$-integral by introducing the density of states at the Fermi level $\nu_F$, which is defined in the normal phase of the superconductor. By carrying out the integral~\cite{SauProxi,PotterComparison}, we end up with Eqs.~\eqref{eq:RealSelfieSimple} and~\eqref{eq:BroadeningSimple}.

{\color{black}

\section{Self-Energy in the Zero Driving Amplitude Limit}\label{app:AppendixBB}

As a consistency check, at this point we note that when the drive amplitude $\beta_{\rm sm}(\bm{p})$ is zero only the zero-th order Bessel functions contribute to the sums in Eqs.~\eqref{eq:RealSelfieSimple} and~\eqref{eq:BroadeningSimple}. Hence, in the respective parameter regimes that these two quantities are non zero, they become diagonal in Floquet space and take the forms:
\bea
-\widetilde{\Sigma}_{{\rm sm};nn}(\varepsilon,\bm{p})
&=&\Gamma \frac{\varepsilon-n\omega-\Delta\tau_1}
{\sqrt{\Delta^2-(\varepsilon-n\omega)^2}}\,,\\
\Gamma_{{\rm sm};nn}(\varepsilon,\bm{p})&=&
\Gamma
\frac{\varepsilon-n\omega-\Delta\tau_1}
{\sqrt{(\varepsilon-n\omega)^2-\Delta^2}}\,{\rm sgn}(\varepsilon-n\omega).\qquad
\label{eq:BareBroad}
\eea

\noi It is important to observe that for $\Delta^2\gg(\varepsilon-n\omega)^2$ one obtains $\widetilde{\Sigma}_{{\rm sm};nn}(\varepsilon,\bm{p})
\rightarrow\Gamma\tau_1$, with $\Gamma$ here functioning as the proximity-induced pai\-ring gap on the semiconductor. In contrast, when $(\varepsilon-n\omega)^2\gg\Delta^2$, one finds $\Gamma_{{\rm sm};nn}(\varepsilon,\bm{p})
\rightarrow\Gamma$, which defines the level broa\-de\-ning acquired by the semiconductor due to its coupling to the metallic bands of the superconductor.

\section{High-Frequency Approximation}\label{app:AppendixDD}

In the high-frequency regime, the condition $\Delta^2>(\varepsilon-\ell\omega)^2$ is satisfied only for the $\ell=0$ mode, while the complementary condition $\Delta^2<(\varepsilon-\ell\omega)^2$ is satisfied by all $\ell\neq0$ modes. Hence, Eqs.~\eqref{eq:RealSelfieSimple} and~\eqref{eq:BroadeningSimple} now become:
\bea
-\widetilde{\Sigma}_{{\rm sm};nm}(\varepsilon,\bm{p})
&=&\Gamma \,e^{i\zeta(n-m)}\,J_n[\beta_{\rm sm}(\bm{p})]\no\\
&\times&\frac{\varepsilon-\Delta\tau_1}
{\sqrt{\Delta^2-\varepsilon^2}}J_m^\dag[\beta_{\rm sm}(\bm{p})]
\,,\label{eq:RealSelfieSimpleHigh}\\
\Gamma_{{\rm sm};nm}(\varepsilon,\bm{p})&\simeq&
\Gamma\,e^{i\zeta(n-m)}\,\sum_{\ell\neq0} \no\\
&\times&J_{n-\ell}[\beta_{\rm sm}(\bm{p})]J_{m-\ell}^\dag[\beta_{\rm sm}(\bm{p})]\,,\quad
\label{eq:BroadeningSimpleHigh}
\eea

\noi where for the evaluation of the matrix elements of the broa\-de\-ning we took into account that $\omega\gg \Delta>|\varepsilon|$.

In the event that the drive amplitude is also much smaller than the driving frequency, i.e., $|\beta_{\rm sm}(\bm{p})|\ll1$, the diagonal elements are given by the approximate forms:
\bea
-\widetilde{\Sigma}_{{\rm sm};nn}(\varepsilon,\bm{p})
&\simeq&\Gamma \frac{\varepsilon-\Delta\tau_1}
{\sqrt{\Delta^2-\varepsilon^2}}\,,\quad\,\,\,{\rm for}\quad n=0,\label{eq:RealSelfieSimpleHighAmpLow}\\
\Gamma_{{\rm sm};nn}(\varepsilon,\bm{p})&\simeq&
\Gamma\,,\qquad\phd\qquad\quad{\rm for}\quad n\neq0\,.\qquad
\label{eq:BroadeningSimpleHighAmpLow}
\eea

\noi In this limit, we can also discard the off-diagonal broa\-dening terms, i.e., $\Gamma_{{\rm sm};n\neq m}(\varepsilon,\bm{p})\simeq0$. However, we leave $\widetilde{\Sigma}_{{\rm sm};n\neq m}(\varepsilon,\bm{p})$ intact for reasons that will become clear later on. In this limit of combined high-frequency and low-amplitude driving, it is straightforward to obtain an effective Hamiltonian for the zero-th Floquet mode. Starting from Eqs.~\eqref{eq:RealSelfieSimpleHighAmpLow} and~\eqref{eq:BroadeningSimpleHighAmpLow}, we jointly obtain the band structure and effective broadening of the zero-th mode by sol\-ving Eq.~\eqref{eq:SemiDynamics} for the retarded Floquet-Green function, i.e., $\sum_m{\cal G}^{R,-1}_{{\rm sm};nm}(\varepsilon,\bm{p})\bm{u}_{\varepsilon,m}(\bm{p})=\bm{0}$, which yields:
\begin{align}
\left[\varepsilon-H_{{\rm sm};0}(\bm{p})+\Gamma\frac{\varepsilon-\Delta\tau_1}{\sqrt{\Delta^2-\varepsilon^2}}\right]\bm{u}_{\varepsilon;0}(\bm{p})=\no\\\sum_{m\neq0}\Lambda_{{\rm sm};0m}(\varepsilon,\bm{p})\bm{u}_{\varepsilon;m}(\bm{p}),
\end{align}

\noi where we introduced the compact notation for $n\neq m$:
\begin{align}
\Lambda_{{\rm sm};nm}(\varepsilon,\bm{p})=H_{{\rm sm};n-m}(\bm{p})+\widetilde{\Sigma}_{{\rm sm};nm}(\varepsilon,\bm{p})\,.
\end{align}

\noi The nonzero $m$-th mode approximately satisfies:
\begin{align}
\big[\varepsilon-m\omega-H_{{\rm sm};0}(\bm{p})+i\Gamma\big]\bm{u}_{\varepsilon;m}(\bm{p})\simeq \Lambda_{{\rm sm};m0}(\varepsilon,\bm{p})\bm{u}_{\varepsilon;0}(\bm{p}),
\end{align}

\noi where we dropped matrix elements of the $m$-th mode to other modes except for the zero-th mode of interest. Within the assumed high-frequency regime we obtain:
\begin{align}
\bm{u}_{\varepsilon;m}(\bm{p})\approx -\frac{1}{(m\omega-i\Gamma)}\Lambda_{{\rm sm};m0}(\varepsilon,\bm{p})\bm{u}_{\varepsilon;0}(\bm{p}).
\end{align}

\noi Plugging the above expression in the equation for the zero-th Floquet mode, we find that the effective inverse Green function for the zero-th Floquet mode discussed in Eq.~\eqref{eq:HighFreqFreenRet}.

}

\section{Dynamical Chiral Symmetry}\label{app:AppendixBold}

To get some insight concerning the symmetries of the driven system in Sec.~\ref{sec:Application}, it is convenient to consider that the superconducting proximity effect solely induces a pairing gap to the semiconductor which is equal to $\Gamma$. Within this oversimplified picture, the nanowire is described by the following toy model Hamiltonian:
\bea
H_{\rm sm}^{\rm toy}(t,p_z)&=&\left(\frac{p_z^2}{2m_{\ast}}-\mu \right)\tau_3+\left[E_Z+\frac{\gamma}{2}\omega\cos(\omega t)\right]\sigma_3\no\\
&&+\alpha p_z\tau_3\sigma_2+\Gamma\tau_1\,.
\label{eq:HamiltonianExampleFiction}
\eea

Understanding the symmetries of the above Hamiltonian is sufficient for inferring the symmetry class of the system, that defines the topological phases which become accessible. First of all, in the absence of the drive, i.e., when $\gamma=0$, the system is characterized by a chiral symmetry which is ge\-ne\-ra\-ted by the operator $\Pi_{\rm static}=\tau_2\sigma_2$, which anticommutes with the Hamiltonian. In addition to this, generalized time-reversal and charge-conjugation symmetries are also present, and are effected by the ope\-ra\-tors $\Theta_{\rm static}={\cal K}$  and $\Xi=\tau_2\sigma_2{\cal K}$, respectively. Here, ${\cal K}$ denotes complex conjugation. The undriven Hamiltonian belongs to symmetry class BDI, which in one spatial dimension supports an integer topological in\-va\-riant. When the driving is switched on, the Hamiltonian preserves a dynamical chiral symmetry~\cite{Yao2017,Kennes,Assili2024}, and remains in BDI symmetry class. This is reflected in the following symmetry relation $
\Pi_{\rm static}^\dag H_{\rm sm}^{\rm toy}(t,p_z)\Pi_{\rm static}=-H_{\rm sm}^{\rm toy}(-t,p_z)$.

\section{High-Frequency Approximation for the Driven Nanowire Hybrid Example}\label{app:AppendixCold}

By employing the results of Sec.~\ref{sec:HFexpansion}, we now obtain the high-frequency expansion for the model of Sec.~\ref{sec:Application}. Spe\-ci\-fi\-cal\-ly, for the model in the latter section we have:
\bea
&&H_{{\rm sm};0}(p_z)= \left(\frac{p_z^2}{2m_{\ast}}-\mu\right)\tau_3+E_Z \sigma_3+\tilde{\alpha}  p_z\tau_3\sigma_2,\\
&&H_{{\rm sm};m\neq0}(p_z)
=\alpha p_z\tau_3 \big[J_m^+(\gamma)\sigma_2-iJ_m^-(\gamma)\sigma_1\big]\no\\
&&\qquad\qquad\quad\,\,\,\,\,\equiv\sigma_3^mJ_m(\gamma)\alpha p_z\tau_3 \sigma_2,\\
&&\widetilde{\Sigma}_{{\rm sm};m0}(\varepsilon,p_z)
=\sigma_3^mJ_0(\gamma/2)J_m(\gamma/2)
\widetilde{\Sigma}_{{\rm sm};00}(\varepsilon,p_z).\qquad
\eea

\noi In the above we introduced the renormalized Rashba SOC strength $\tilde{\alpha}=\alpha J_0(\gamma)$, while $\widetilde{\Sigma}_{{\rm sm};00}(\varepsilon,p_z)
$ is defined by the expression in Eq.~\eqref{eq:RealSelfieSimpleHighAmpLow}. Given the above, we find that for any nonzero value of $m$ we have:
\bea
\Lambda_{{\rm sm};m0}(\varepsilon,p_z)
&=&\sigma_3^m\Big[J_0(\gamma/2)J_m(\gamma/2)
\widetilde{\Sigma}_{{\rm sm};00}(\varepsilon,p_z)\no\\
&&\quad\,+J_m(\gamma)\alpha p_z\tau_3 \sigma_2\Big]\,.\eea

To infer the band renormalization and broadening effects, it is required to calculate the following product:
\bea
&&\Lambda_{{\rm sm};0m}(\varepsilon,p_z)\Lambda_{{\rm sm};m0}(\varepsilon,p_z)\equiv\Lambda_{{\rm sm};m0}^\dag(\varepsilon,p_z)\Lambda_{{\rm sm};m0}(\varepsilon,p_z)=\no\\
&&=\big[J_0(\gamma/2)J_m(\gamma/2)\widetilde{\Sigma}_{{\rm sm};00}(\varepsilon,p_z)+J_m(\gamma)\alpha p_z\tau_3\sigma_2\big]^2\no\\
&&=\big[J_0(\gamma/2)J_m(\gamma/2)\widetilde{\Sigma}_{{\rm sm};00}(\varepsilon,p_z)\big]^2+\big[J_m(\gamma)\alpha p_z\big]^2\no\\
&&\,\quad-2J_0(\gamma/2)J_m(\gamma/2)J_m(\gamma)\,\frac{\Gamma\varepsilon}{\sqrt{\Delta^2-\varepsilon^2}}\,\alpha p_z\tau_3\sigma_2\,.\qquad\eea

\noi Note that since the above expression is even under $m\mapsto-m$, the respective sum in the second row of Eq.~\eqref{eq:HighFreqFreenRet} does not contribute, thus, no energy renormalization arises. Therefore, in the high-frequency regime, the zero-th Floquet mode solely experiences the renormalization of the Rashba SOC strength along with level broa\-de\-ning. Using the above expressions, it is now straightforward to obtain the effective broa\-dening of the zero-th Floquet mode, which is defined as $\Gamma_0(\varepsilon,p_z)=\Gamma\sum_{m\neq0}\,\Lambda_{{\rm sm};m0}^\dag(\varepsilon,p_z)\Lambda_{{\rm sm};m0}(\varepsilon,p_z)/\big[(m\omega)^2+\Gamma^2\big]$.

Concluding this appendix, we remark that for large driving frequencies and weak amplitudes, we can consider that the main contribution to the broadening stems only from $m=\pm1$, in which case we obtain:
\begin{align}
\Gamma_0(\varepsilon,p_z)\approx\frac{\Gamma}{2}\left(\frac{\gamma}{\omega}\right)^2\Big[\frac{1}{2}\widetilde{\Sigma}_{{\rm sm};00}(\varepsilon,p_z)+\alpha p_z\tau_3\sigma_2\Big]^2.\end{align}

\noi which is consistent with the expected form of a perturbative result at second order with respect to $\gamma$.

\section{Topological Invariants with Broadening}\label{app:AppendixDold}

Since in the high-frequency regime the system behaves essentially as a static system, we can employ the above fin\-dings and evaluate the topological invariant using the Euclidean (imaginary time) causal Green function, which is obtained by means of an analytical continuation from the retarded Green fuction~\cite{Mahan}. Hence, transferring to ima\-gi\-na\-ry time $\varepsilon\mapsto i\epsilon$, the analytical continuation yields the Euclidean Green function:
\bea
&&{\cal G}^{-1}_{{\rm sm}}(\epsilon,p_z)=i\epsilon\left(1+\frac{\Gamma}{\sqrt{\Delta^2+\epsilon^2}}\right)- \left(\frac{p_z^2}{2m_{\ast}}-\mu\right)\tau_3\no\\
&&-E_Z \sigma_3-\tilde{\alpha} p_z\tau_3\sigma_2-\frac{\Gamma\Delta\tau_1}{\sqrt{\Delta^2+\epsilon^2}}+i{\rm sgn}(\epsilon)\Gamma_0(p_z),\qquad
\eea

\noi where, with no loss of generality, we discarded the ener\-gy dependence of the broadening. It is well established that the topological invariant associated with the above Green function can be obtained from the Green function by setting $\epsilon=0$. See for instance the works in Refs.~\onlinecite{Volovik,SilaevVolovik,Gurarie,ZhongWang}. This imme\-dia\-te\-ly shows that the broadening does not affect the value of the topological invariant. Hence, the topological phase diagram can be directly obtained by defining the effective Hamiltonian at zero quasi-energy:
\bea
{\cal H}_{\rm sm}(p_z)\equiv-{\cal G}^{-1}_{{\rm sm}}(\epsilon=0,p_z)&=& \left(\frac{p_z^2}{2m_{\ast}}-\mu\right)\tau_3+E_Z\sigma_3\no\\
&&+\tilde{\alpha} p_z\tau_3\sigma_2+\Gamma\tau_1.
\label{eq:high-fr-regime}
\eea

\noi Since topological phase transitions take place by means of gap closings and reopenings at $p_z=0$, we find that in the high-frequency and low-amplitude limit the driving cannot alter the topological properties of the system.

\end{document}